\documentclass[useAMS,usenatbib]{mn2e}
\usepackage[totalwidth=515pt,totalheight=680pt,left=1.4cm,right=1.4cm]{geometry}
\usepackage{graphicx,amssymb}
\input psfig

\def\epsilonW{{a_h}}
\def\OmegaG{{\Omega_{\rm G}}}
\def\Rd{{R_{\rm d}}}
\def\Rc{{R_{\rm c}}}
\def\Mb{{M_{\rm b}}}
\def\vc{{v_{\rm c}}}

\title[Post-MS evolution in a Galactic context]
{The Great Escape III: Placing post-main-sequence evolution
of planetary and binary systems in a Galactic context}
\author[Veras, Evans, Wyatt \& Tout]
{Dimitri Veras$^{1,2}$\thanks{E-mail: d.veras@warwick.ac.uk},
N. Wyn Evans$^{2}$,
Mark C. Wyatt$^{2}$ and
Christopher A. Tout$^{2}$ 
\\
$^{1}$Department of Physics, University of Warwick, Coventry CV4 7AL
\\
$^{2}$Institute of Astronomy, University of Cambridge, Cambridge CB3 0HA}

\begin{document}

\date{
Accepted 2013 October 4.  Received 2013 September 26; in original form 2013 July 23}

\pagerange{\pageref{firstpage}--\pageref{lastpage}} \pubyear{2014}

\maketitle

\label{firstpage}

\begin{abstract}
Our improving understanding of the life cycle of planetary systems
prompts investigations of the role of the Galactic environment before,
during and after Asymptotic Giant Branch (AGB) stellar evolution.  Here, we investigate
the interplay between stellar mass loss, Galactic tidal perturbations,
and stellar flybys for evolving stars which host one planet, smaller
body or stellar binary companion and reside in the Milky Way's bulge
or disc.  We find that the potential evolutionary pathways from a main 
sequence (MS) to a white dwarf (WD) planetary system are a strong
function of Galactocentric distance only with respect to the
prevalence of stellar flybys.   Planetary ejection and collision
with the parent star should be more common towards the bulge.
At a given location anywhere in the Galaxy, if the mass loss is adiabatic, then the
secondary is likely to avoid close flybys during AGB evolution, and
cannot eventually escape the resulting WD because of Galactic tides
alone.  Partly because AGB mass loss will shrink a planetary system's
Hill ellipsoid axes by about 20 to 40 per cent, Oort clouds orbiting WDs
are likely to be more depleted and dynamically excited than on the MS.
\end{abstract}

\begin{keywords}
planets and satellites: dynamical evolution and stability --
planet-star interactions -- stars: AGB and post-AGB -- stars:
evolution -- The Galaxy: kinematics and dynamics -- Oort Cloud
\end{keywords}

\section{Introduction}

Planetary systems do not evolve in isolation.  Their nascent orbital
architectures may have been moulded by irradiation from, and the
mutual interactions amongst, stars in a stellar birth cluster
\citep{lauada1998,bonetal2001,smibon2001,adaetal2006,
freetal2006,maletal2007,pargoo2009,spuetal2009,thietal2011,deetal2012,parqua2012,marpic2013,thompson2013}.
Later, on the main sequence (MS), planetary orbits may be further
disrupted by flybys from Galactic disc stars \citep{zaktre2004,
maletal2011,boletal2012,vermoe2012}. The effect of
flybys on debris discs and potential system habitability may be a strong 
function of Galactic environment \citep{jimetal2011,jimetal2013}.
Gravitational scattering amongst multiple planets may be induced due to the effect of Galactic 
tides on wide binary stellar companions \citep{kaietal2013}.  Oort clouds 
continuously interact with the Galactic environment; their content and 
extent are dictated by the combination of global Galactic tides as well as
stellar flybys
\citep{heitre1986,dunetal1987,dybczynski2006,emeetal2007,ricetal2008,kaiqui2009,braetal2010,colsar2010,bramor2013,rickman2013}.

Modelling these effects during post-MS evolution provides new and
largely unexplored challenges, and is the focus of this study.  When
stars evolve off the MS, they shed mass and expand their envelopes,
occasionally out to several au.  Planetary material too close to the
star tidally interacts with the envelope
\citep{soker1998,villiv2007,caretal2009,villiv2009,noretal2010,kunetal2011,villaver2011,musvil2012,pasetal2012,spiegel2012,adablo2013,norspi2012},
with destruction being the likely, but not guaranteed, outcome
\citep{maxetal2006,beaetal2011,beasok2012}.

A single orbiting body which survives giant branch evolution alters
its orbit in a well-defined, analytically-precise manner
in the isotropic mass loss limit
\citep{omarov1962,hadjidemetriou1963} as well as when the 
mass is lost anisotropically \citep{veretal2013a}.
Detailed non-tidal studies of exoplanetary systems with 
stellar mass loss include post-MS explorations of

\newpage

\begin{itemize}
\item{1-planet, 1-star systems: (\citealt*[][hereinafter
  Paper I]{veretal2011}; \citealt*{adetal2013})  }
\item{1-planet, 2-star systems: (\citealt*{kraper2012}; \citealt*[][hereinafter
  Paper II]{vertou2012})} 
\item{2-planet, 1-star systems: \citep{debsig2002,veretal2013b,voyetal2013}} 
\item{3-planet, 1-star systems: \citep{debsig2002,musetal2013a}} 
\item{2-planet, 2-star systems: (\citealt*{por2013}; \citealt*{musetal2013b})} 
\item{0-planet, 1-star systems with a disc or cloud: \citep{paralc1998,bonwya2010} } 
\item{1-planet, 1-star systems with a disc or cloud: \citep{bonetal2011,debetal2012} } 
\end{itemize}

\noindent{}Also, more generally, the restricted three-body problem with mass loss
has been studied in a broad analytical context 
\citep[e.g.][]{singh2011,varhad2012,zhang2012}.

Here, we compare the effects of mass loss, Galactic tides and stellar
flybys for a single secondary body (e.g. planet, asteroid, comet or
distant non-evolving star) orbiting a single primary star, and place
these perturbations in the context of the entire lifetime of planetary
systems, from the MS to the white dwarf (WD) phases. We consider a
range of possible planetary separations, stellar separations from the
Galactic centre, and WD progenitors.  In Section 2, we describe our
Galactic tidal model and present the equations of motion.  Section 3
quantifies each effect and compares their strengths and reach.  We
give specific properties of the resulting orbital motion and outline
potential evolutionary pathways in Section 4.  Section 5 discusses
implications and extensions, before we make our conclusions in Section
6.

\section{Galactic Model}

We consider a primary star of mass $M_{\star}$ that is orbiting the
centre of the Milky Way Galaxy on a circular orbit in the Galactic plane.
The secondary of mass $M_{\rm p}$ orbits the primary 
on an arbitrarily wide, elliptic and inclined orbit
with respect to a reference plane that is parallel to
and coincident with the Galactic plane.  Both the primary and 
secondary are treated as point masses.  The secondary may represent 
any non-evolving stellar companion; without loss of 
generality, we will often refer to it as a planet.
This planetary system is
orbiting the centre of Galaxy with
circling frequency $\OmegaG$ counterclockwise from the point of view
of the North Galactic pole.  The system is in the Galactic plane at a
distance $R$ from the Galactic centre. 
The star is undergoing post-main
sequence evolution and is losing mass isotropically at the rate
$\dot{M}_{\star}$; the isotropic mass loss approximation is robust
for post-MS systems \citep{veretal2013a}.

\subsection{Equations of motion}

We adopt the same right-handed stellarcentric Cartesian coordinate
system as \cite{heitre1986}, \cite{braetal2010}, and
\cite{vereva2013a,vereva2013b}, with $x$ pointing radially outwards,
$y$ pointing tangent to a star's motion in the Galactic disc and
positive in the direction of Galactic rotation, and $z$ pointing in
the direction of the South Galactic Pole.  The equations of motion for
the planet are then

\begin{equation}
\frac{d^2x}{dt^2} = - \frac{G \left(M_{\star}(t) + M_{\rm p} \right) x}{\left(x^2 + y^2 + z^2\right)^{3/2}} 
+ \Upsilon_{xx} x + \Upsilon_{xy} y
,
\label{xeq}
\end{equation}

\begin{equation}
\frac{d^2y}{dt^2} = - \frac{G \left(M_{\star}(t) + M_{\rm p} \right) y}{\left(x^2 + y^2 + z^2\right)^{3/2}} 
+ \Upsilon_{yx} x + \Upsilon_{yy} y
,
\label{yeq}
\end{equation}

\noindent{and}

\begin{equation}
\frac{d^2z}{dt^2} = - \frac{G \left(M_{\star}(t) + M_{\rm p} \right) z}{\left(x^2 + y^2 + z^2\right)^{3/2}} 
+ \Upsilon_{zz} z
,
\label{zeq}
\end{equation}

\noindent{where} the $\Upsilon$ terms are due to the Galactic tide.
The perturbative acceleration due to isotropic mass loss is implicit
-- a phenomenon explained by \cite{hadjidemetriou1963} -- and hence
does not appear explicitly in equations~(\ref{xeq})-(\ref{zeq}).  The
following form of the $\Upsilon$ terms for circular stellar orbits in
the Galactic disc around the Galactic centre are derived by
\cite{braetal2010} and \cite{vereva2013b}:

\begin{equation}
\Upsilon_{xx} =  \OmegaG^2 [ (1-\delta) \cos{\left(2 \OmegaG t\right)} -\delta ]
,
\label{xx}
\end{equation}

\begin{equation}
\Upsilon_{xy} =  \OmegaG^2 (1 - \delta) \sin{\left(2 \OmegaG t\right)}
,
\label{xy}
\end{equation}

\begin{equation}
\Upsilon_{yx} =  \OmegaG^2 (1 - \delta) \sin{\left(2 \OmegaG t\right)}
,
\label{yx}
\end{equation}

\begin{equation}
\Upsilon_{yy} =  -\OmegaG^2 [ (1 -\delta)\cos{\left(2 \OmegaG t\right)} + \delta]
,
\label{yy}
\end{equation}

\noindent{and}

\begin{equation}
\Upsilon_{zz} =  -\left[ 4 \pi G \rho_{\rm tot} -2 \delta \OmegaG^2 \right]
,
\label{zz}
\end{equation}

\noindent{where} $\delta \equiv -(A+B)/(A-B)$ represents the
logarithmic gradient of the Galactic rotation curve in terms of the
Oort Constants $A$ and $B$~\citep{matwhi1996}.  The expressions for
$\OmegaG$, $A$ and $B$ are obtained from the
three-component Galaxy model presented by \cite{vereva2013b},
and the total Galactic density $\rho_{\rm tot}$ is described below.

\subsection{Matter density}

The total density of the Galaxy, including both stars and dark matter,
is denoted by $\rho_{\rm tot}(R,z)$ in terms of cylindrical polar
coordinates $(R,z$).  This density is composed of an exponential disc, a
Hernquist bulge and a cored isothermal halo that together reproduce
the local stellar kinematics.

The stellar bulge has a Hernquist profile with mass $\Mb$ and scalelength $\epsilonW$:
\begin{equation}
\rho_{\rm bulge}(R,z) = \frac{\epsilonW \Mb}{2 \pi \sqrt{R^2 + z^2} 
\left(\sqrt{R^2 + z^2} + \epsilonW \right)^3},
\label{bulgedensity}
\end{equation}
The stellar disc is a double exponential with scalelength $\Rd$ and
scaleheight $h$:
\begin{equation}
\rho_{\rm disc}(R,z) = \frac{\Sigma_0}{2h} \exp{\left( -\frac{R}{\Rd} \right)}
                  \exp{\left(-\frac{\left|z\right|}{h}\right)},
\label{discdensity}
\end{equation}
The dark matter halo is a cored, isothermal model with core radius
$\Rc$ and asymptotic amplitude of rotation curve $v_0$
~\citep{evans1993}:
\begin{equation}
\rho_{\rm halo}(R,z) =      {v_{0}^2 \over 4 \pi G q^2}{
\Rc^2 \left(1 + 2q^2 \right) + R^2 + z^2(2-q^{-2}) \over
\Rc^2 + R^2 + z^2q^{-2} }.
\label{halodensity}
\end{equation}
We adopt the same numerical values as \cite{vereva2013b} for the
bulge mass, $\Mb = 3.6 \times 10^{10} M_{\odot}$, Hernquist potential
parameter $\epsilonW = 0.7$ kpc, thin disc scale height $h = 0.3$ kpc,
thin disc scale length $\Rd = 3$ kpc, circular velocity $v_0 = 215$ 
km s$^{-1}$, core radius $\Rc = 16$ kpc, normalisation surface density
constant $\Sigma_0 = 51 \times \exp{\left(8{\ \rm kpc}/\Rd \right) }
M_{\odot}{\rm pc}^{-2}$ and equipotential axis ratio $q =1$.
Adopting these values for our model allows us to closely reproduce 
observations of all the local stellar kinematics, as described
in \cite{vereva2013b}.

The Galaxy model that we use reproduces the local stellar
kinematics. It has of course some arbitrariness, as the
three-dimensional density distributions of the bulge and halo are not
well known.  For example, cusped Navarro-Frenk-White (NFW) models are
often used for halos, motivated by dissipationless cosmological
simulations. They differ from cored isothermal halos, particularly at
the centre. However, when baryonic effects are included in galaxy
formation simulations, they transfer energy between the luminous and
dark components, producing a shallower or cored profile for the dark
matter. This is consistent with observational data on the Galaxy,
which precludes strongly cuspy halos (like the NFW) using constraints
from the rotation curve and the microlensing experiments~\citep{BE}.
Similarly, there are a number of possible models for the Galactic
bulge, depending on whether its origin is through mergers or through
secular disk processes or though thickening of a Galactic
bar~\citep{KK}.  We have used a Hernquist model, but exponential or
Sersic models could easily have been used instead. The main advantage
of a Hernquist model is that the gravitational force-field is
analytic, which speeds numerical orbit integrations. Sersic models in
general do not have analytic potentials.

\subsection{Frequencies and velocities}

The circular frequency in the Galactic plane is given by
\[
\OmegaG^2(R) = \underbrace{\frac{G M_b}{R\left(R + \epsilonW\right)^2}}_{\rm bulge} 
             + \underbrace{\frac{v_{0}^2}{\Rc^2+R^2}}_{\rm halo} +
\nonumber
\]
\begin{equation}
\underbrace{\frac{G \pi \Sigma_0}{\Rd}
\left[  
I_0\left( \frac{R}{2\Rd} \right)
  K_0\left( \frac{R}{2\Rd} \right) \!-\!I_1\left( \frac{R}{2\Rd}
  \right) K_1\left( \frac{R}{2\Rd} \right)
\right]
}_{\rm disc}
,
\label{OmG}
\end{equation}
where $I_0$, $I_1$, $K_0$ and $K_1$ are modified Bessel functions.
The circular velocity $\vc = R \OmegaG$.   The Oort constants are

\begin{equation}
A = \frac{1}{2}
\left[  
\OmegaG -
\frac{d\vc}{dR}
\right], 
\qquad
{\rm and}
\qquad
B = - \frac{1}{2}
\left[  
\OmegaG +
\frac{d\vc}{dR}
\right]
,
\end{equation}

\noindent{with}

\[
\frac{d\vc}{dR} =  
\underbrace{\frac{G \Mb\left(\epsilonW - R\right)}{2 \OmegaG R\left(R + \epsilonW\right)^3}}_{\rm bulge}
+
\underbrace{\frac{\Rc^2v_{0}^2}{\OmegaG \left(\Rc^2 + R^2\right)^2}}_{\rm halo}
\]

\[
+ \frac{G \pi \Sigma_0}{2 \OmegaG \Rd^2}
\bigg\lbrace
R I_1 \left( \frac{R}{2\Rd} \right) K_0 \left( \frac{R}{2\Rd} \right) 
\]

\begin{equation}
+ I_0 \left( \frac{R}{2\Rd} \right)
\left[2 \Rd K_0 \left( \frac{R}{2\Rd} \right) - R K_1 \left( \frac{R}{2\Rd} \right)  \right]
\bigg\rbrace
\label{dvcdr}.
\end{equation}
The last term is the contribution from the Galactic disc.  This term has
been simplified by exploiting the relationships between Bessel functions of
different orders.

\subsection{Re-expressed perturbative accelerations}

Equations~(\ref{bulgedensity})-(\ref{dvcdr}) allow us to re-express
the form of the perturbations in equations~(\ref{xx})-(\ref{zz}) and
eliminate $\delta$.  Doing so will aid further analysis.  First, note that
$\OmegaG \delta = d\vc/dR$ and $\left(1 - \delta \right) \OmegaG =
\OmegaG - d\vc/dR$.  Consequently, equations~(\ref{xx})-(\ref{zz}) can
be rewritten as
\begin{equation}
\Upsilon_{xx} = 
-\OmegaG \frac{d\vc}{dR} - ( \OmegaG^2 - \OmegaG \frac{d\vc}{dR} )
\cos{(2 \OmegaG t)}
,
\label{xxnew}
\end{equation}
\begin{equation}
\Upsilon_{yy} = 
-\OmegaG \frac{d\vc}{dR} + ( \OmegaG^2 - \OmegaG \frac{d\vc}{dR} ) \cos{(2 \OmegaG t)},
\label{yynew}
\end{equation}
\begin{equation}
\Upsilon_{xy} = \Upsilon_{yx} =  -\left( \OmegaG^2 -  \OmegaG \frac{d\vc}{dR} \right) \sin{\left(2 \Omega_G t\right)}
\label{xynew}
\end{equation}
\noindent{and}

\[
\Upsilon_{zz} =  -\left[ 4 \pi G \rho_G -2 \OmegaG \frac{d\vc}{dR} \right]
             =  - \OmegaG^2|_{\rm bulge}
                - \OmegaG^2|_{\rm halo}
\]

\[
  - 
G \pi \Sigma_0 \bigg[\frac{2}{h} \exp{\left(-\frac{R}{\Rd}\right)} -  
\frac{1}{\Rd^2}  
\bigg\lbrace
R I_1 \left( \frac{R}{2\Rd} \right) K_0 \left( \frac{R}{2\Rd} \right)
\]

\begin{equation}
+ I_0 \left( \frac{R}{2\Rd} \right)
\bigg[2 \Rd K_0 \left( \frac{R}{2\Rd} \right) - R K_1 \left( \frac{R}{2\Rd} \right)  \bigg]
\bigg\rbrace \bigg]
,
\label{zznew}
\end{equation}

\noindent{where} the last vertical perturbative term is the contribution
from the disc.  We will later show that the quantity in parenthesis for the 
planar cross-term (in equation \ref{xynew}) scales with the region of gravitational
influence of an individual star.  Also, note that 
the bulge and halo components of equation (18) 
reduce exactly to the corresponding components of equation~(\ref{OmG}).

\begin{figure*}
\centerline{
\psfig{figure=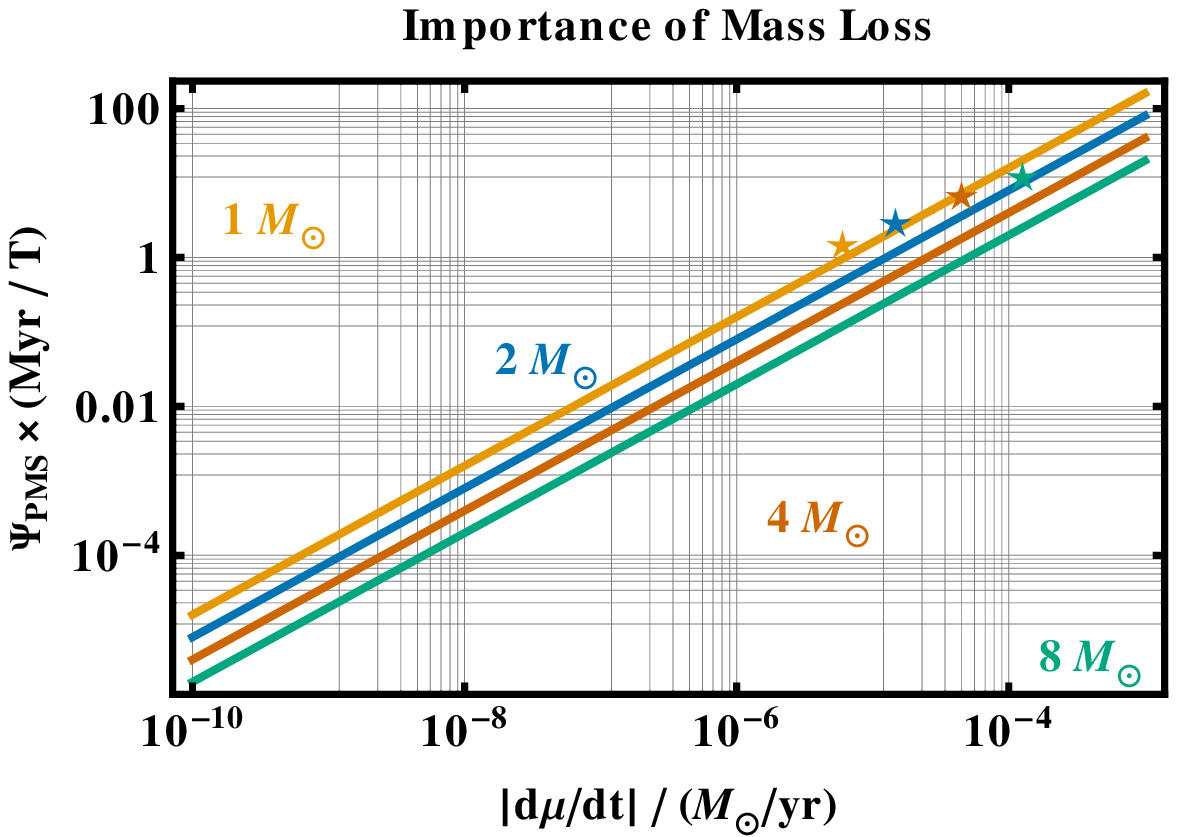,height=7.0cm,width=9.0cm} 
\psfig{figure=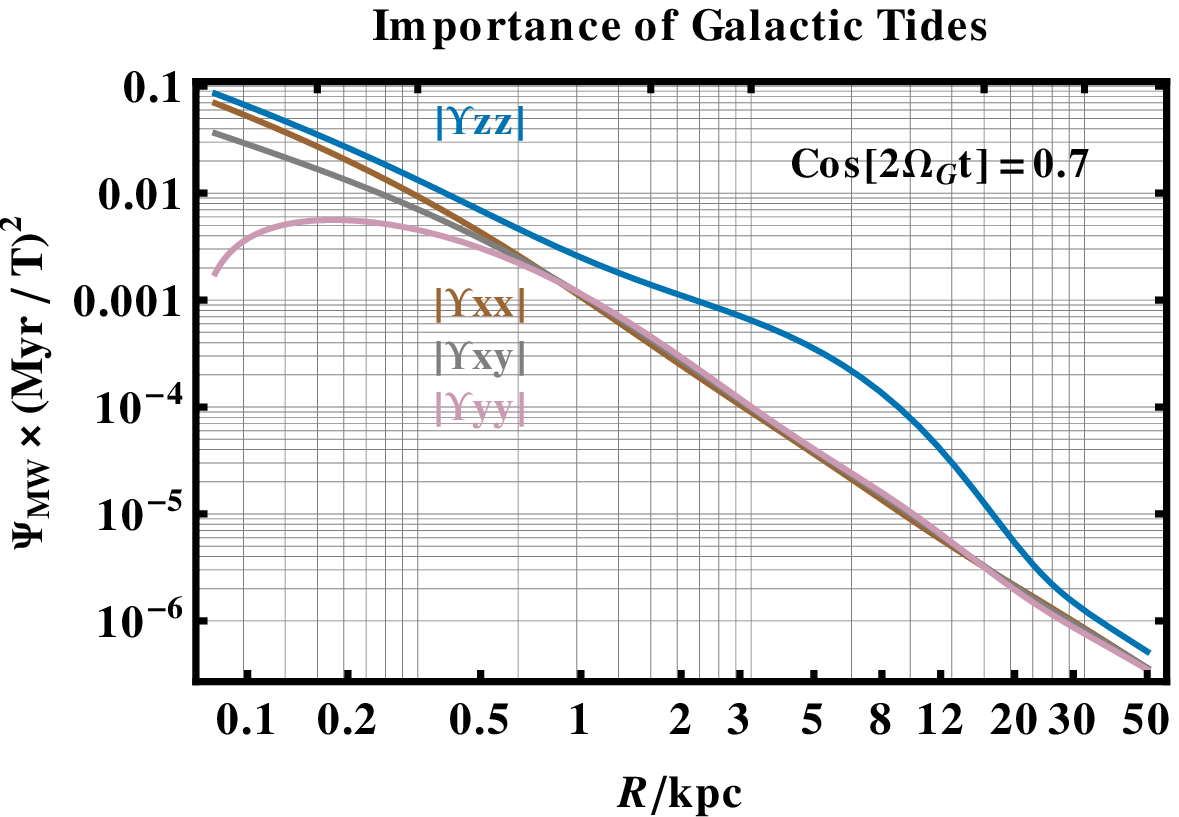,height=7.0cm,width=9.0cm}
}
\caption{A demonstration that orbital effects due to post-MS mass loss may 
be decoupled from the effects due to Galactic tides.
The dimensionless adiabaticity indices ($\Psi_{\rm PMS}, \Psi_{\rm MW}$) for
stellar mass loss ({\it left panel}) and for a snapshot of Galactic
tides ({\it right panel}) represent how actively a secondary's
orbit is influenced by these effects.
Nonadiabatic influences typically become important when $\Psi \gtrsim 0.02$.
$T$ represents the orbital period of the
secondary, and five-pointed stars indicate the highest
$\Psi_{\rm PMS}$ reached in stellar evolution
models. 
Although the Galactic tide adiabaticity index is a function of time,
the maximum perturbation is well-represented by the topmost curve
for all times and Galactocentric distances.  }
\label{PMSIndex}
\end{figure*}

\section{Comparison of effects}

Having presented our Galactic model, we now characterize the regimes
in which different effects may be important by comparing the
timescales for orbital motion, mass loss, and Galactic tides.  We then
place impulsive stellar encounters in this picture.  We conclude with
plots that combine all these effects.

\subsection{Post-MS timescale}

We define two nondimensional indices.  The first is from Paper I:

\[
\Psi_{\rm PMS}(t) \equiv \frac{\rm orbital \ timescale}{\rm mass-loss \ timescale}
= \frac{\dot{\mu}(t)}{n(t)\mu(t)}
\]
~
\begin{equation}
\approx 0.16
\left( \frac{\dot{M_{\star}}(t)}{1 M_{\odot} {\rm yr}^{-1}}\right)
\left( \frac{a(t)}{1 {\rm AU}}\right)^{\frac{3}{2}}
\left( \frac{M_{\star}(t) + M_{\rm p}}{1 M_{\odot}}\right)^{-\frac{3}{2}}
,
\label{PMSindex}
\end{equation}

\noindent
where $a$ represents the planet's semimajor axis, $n$ the planet's
mean motion and $\mu = G \left(M_{\star} + M_{\rm p}\right)$.

The index $\Psi_{\rm PMS}$ provides a measure of the mass-loss
adiabaticity of a system.  In a system which is adiabatic with
respect to mass loss, the semimajor
axis increases at a steady rate and the eccentricity remains
unchanged\footnote{There is a typographic error in equation~(17) of
  Paper I; the denominator should read 
$1 + e_0 \cos{f}$ (G. Voyatzis, priv. comm).}.  In a
nonadiabatic system, the eccentricity can change and trigger instability.
The transition is not sharp; substantial deviations from adiabaticity
can occur at $\Psi_{\rm PMS} \approx 0.02$ (see e.g. fig.~3 of Paper
I).  

In Fig.~\ref{PMSIndex}, we plot $\Psi_{\rm PMS}$ as a function of mass-loss 
rate for four different stellar masses.  $\Psi_{\rm PMS}$ is scaled
by the secondary orbital period $T$.  The coloured stars
indicate the highest $\Psi_{\rm PMS}$ values achieved during simulations of
stellar evolutionary models with the {\tt SSE} code
\citep{huretal2000}.  The {\tt SSE} code uses a Reimers mass 
loss prescription with a coefficient of $0.5$ on the Red Giant Branch 
(RGB) and the mass loss prescription from \cite{vaswoo1993} on the Asymptotic 
Giant Branch (AGB).  The stars in the figure do not lie on the lines because the highest value of
$\Psi_{\rm PMS}$ is achieved in each case after the star has
lost some of its original mass.  The highest values of 
$\Psi_{\rm PMS}$ achieved in the simulations correlates
closely to the highest values of mass loss experienced
by the parent stars.  The figure demonstrates, for example,
that a planet orbiting a $1M_{\star}$ star at $a = 10^4$ au 
experiences nonadiabatic evolution if the parent star loses mass at a
rate greater than about $10^{-7} M_{\odot}$ yr$^{-1}$.  As the position of
the yellow star suggests, this mass-loss rate is thought to be easily 
achieved by AGB stars.

\subsection{Galactic tidal timescale}

We define the Galactic tidal timescale index as:

\begin{equation}
\Psi_{\rm MW}(t) \equiv \frac{\rm orbital \ timescale}{\rm Galactic \ tidal \ timescale}
             \approx \frac{\Upsilon(t)}{n(t)^2}.
\label{PsiMW}
\end{equation}

\noindent
This index provides a measure of the Galactic tidal
adiabaticity of a system.  The orbital properties
of adiabatic systems measured with respect to Galactic tides
are different than those measured with respect to mass
loss, and will be described later in Sections 4.1-4.2.
The time dependence of $\Upsilon$ arises from the planar tidal terms
only.  However, as we will show max$\left|\Upsilon\right| =
\left|\Upsilon_{zz}\right|$ everywhere except within the inner
$500$ pc area of the Milky Way.  Further, in this inner bulge
region, $\left|\Upsilon_{zz}\right|$ is comparable to the maximum
planar perturbation.  Hence, we approximate $\left|\Upsilon\right|$ by
$\left|\Upsilon_{zz}\right|$ only.

In the right panel of Fig.~\ref{PMSIndex}, we plot
$\left|\Upsilon_{zz}\right|$, as well as 
a snapshot of $\left|\Upsilon_{xx}\right|$
and $\left|\Upsilon_{xy}\right|$, which are time-dependent.  The
approximate maximum value of $\left|\Upsilon\right|$ for all times and
Galactocentric distances is the topmost curve.  The plot
demonstrates that the prospects for a companion to undergo
nonadiabatic evolution due to Galactic tides is a strong function of
both Galactic position and orbital period.  Assuming an adiabatic limit
of $\Psi_{\rm MW} \approx 0.02$, a planet with a period of
1 Myr (corresponding to $a = 10^4$ AU around a $1M_{\odot}$ star) 
is affected by tides adiabatically throughout the Milky Way except
within the first few hundred parsecs.  However, increasing this
semimajor axis by a factor of 5 lowers the adiabaticity boundary 
by a factor of 125, causing any planet within the Solar
circle ($R \lesssim 8$ kpc) to undergo nonadiabatic evolution due to
Galactic tides.

\subsection{Timescale comparison}

Comparing both plots in Fig.~\ref{PMSIndex} side-by-side helps
us to establish the expected level of coupling between mass
loss and Galactic tides with respect to orbital excitation.
For a given planetary orbital period (fixed $T$), and for any parent 
star experiencing post-MS evolution, $\Psi_{\rm PMS} \gtrsim \Psi_{\rm MW}$
holds true, and often $\Psi_{\rm PMS} \gg \Psi_{\rm MW}$ is satisfied.  Hence,
the effects of post-MS mass loss and Galactic tides can effectively
be decoupled when the former is active.

\subsection{The Hill ellipsoid}

If the secondary is sufficiently far from the primary, then 
the secondary will
escape. \cite{vereva2013b} showed that the Hill sphere in this
instance is actually a triaxial ellipsoid with extent
\begin{equation}
\vec{r}_{\rm Hill} = \left( \frac{GM_{\star}}{\alpha} \right)^{1/3} \vec{k}
\label{HillShrink}
\end{equation}
such that
\begin{equation}
\vec{k} = \left( 1, \frac{2}{3}, 
\frac{\left[Q \left( 1 + \sqrt{1+ Q} \right)  \right]^{2/3} - Q}
{\left[Q \left(1 + \sqrt{1+Q}\right) \right]^{1/3}}  
\right)
,
\label{HillShrink2}
\end{equation}
where $Q \equiv -\alpha/\Upsilon_{zz}$ and\footnote{There is a sign
  error in the definitions of $\alpha$ and $Q$ in \cite{vereva2013b}.}
$\alpha \equiv 4A\left(A-B\right)$.  The third ($z$) component of
$\vec{k}$ lies in the range $\left(0, 2/3\right)$.  Simplification
shows that $\alpha = 2\left[ \OmegaG^2 - \OmegaG \left( d\vc/dR
  \right)\right]$, meaning that the perturbations of 
equations~(\ref{xxnew}-\ref{xynew}) can be expressed simply in 
terms of the Hill ellipsoid axes through 
$\Upsilon_{xx} = -\OmegaG^2 + (\alpha/2) [1 - \cos(2\OmegaG t)]$,
$\Upsilon_{yy} = -\OmegaG^2 + (\alpha/2) [1 + \cos(2\OmegaG t)]$
and $\Upsilon_{xy} = \Upsilon_{yx} = -2\alpha \sin{\left(2
  \OmegaG t \right)}$.  Therefore, expressed in Hill radii, planar
perturbation maxima are

\begin{equation}
\left|\Upsilon_{xy}\right|_{\rm max} = 
\frac{2GM_{\star}}{r_{{\rm Hill},x}^3}
\end{equation}

\noindent{and}

\begin{equation}
\left|\Upsilon_{xx}\right|_{\rm max} = \left|\Upsilon_{yy}\right|_{\rm max} = 
\frac{GM_{\star}}{r_{{\rm Hill},x}^3} - \OmegaG^2.
\end{equation}

Having defined the Hill ellipsoid and related it to the magnitude
of the perturbations, we can now assess the adiabaticity of orbits
at the Hill limit.  Using equation (\ref{HillShrink}), we find that
the slowest possible mean motion of a companion is
\begin{equation}
n_{\rm min} =
\left[\frac{1 - e^2}{1 + e \cos{f}}  \right]^{\frac{3}{2}}
\sqrt{\left| \frac{\alpha}{k_{z}^3} \right|}
< \frac{3}{2} \sqrt{\frac{3 \alpha}{2}}
\left(1 + e\right)^{\frac{3}{2}}
,
\label{minn}
\end{equation}
\noindent{where} $e$ is eccentricity and $f$ is the true anomaly.
Further, in the limit of small $Q$ and for circular orbits, $n_{\rm
  min} \approx \sqrt{\Upsilon_{zz}/2}$.  Hence, we can combine this
relation with equation~(\ref{PsiMW}) in order to relate the minimum
Hill semiaxis ($z$) with the adiabatic Galactic tidal limit.  Doing so
gives $T_{\rm Hill}/T_{\rm Gal} \approx \sqrt{2/\Psi_{\rm MW}}$,
independent of $R$, where $T_{\rm Hill}$ and $T_{\rm Gal}$ represent
the orbital periods at the Hill limit and the nonadiabatic Galactic
limit.  Therefore, Galactic effects become non-adiabatic at a nearly fixed 
fraction of the Hill limit, and orbits at the Hill limit are always non-adiabatic.

\subsection{Stellar flybys}

A final consideration is the effect of stellar flybys on planetary systems.  
This type of perturbation is a scaled-down version of the well-studied and more 
general scenario of stars flying past binary stellar systems 
\cite[Chapters 6,8,10 of][]{valkar2006}.  The seminal work of 
\cite{heggie1975} helped lay the analytical framework for modeling this dynamical 
three-body interaction.  Early numerical integration ensembles 
\citep{hilful1980,hills1984a} provided further insight that is sometimes difficult 
to parse from the analytics.  Subsequent investigations focusing on the Solar 
System \citep[e.g.][]{hills1984b,hills1986} linked flybys with the Oort cloud, 
and others studies considered repeated flybys in different astrophysical contexts.  
\cite{hills1984c}, for example, found that binaries in globular clusters are 
more resistant to changes in semimajor axis than eccentricity.

Such orbital element variation tendencies will have important consequences for detailed 
studies of repeated flybys on planetary systems.  \cite{hegras1996} significantly 
facilitated understanding of how orbits change due to flybys by deriving explicit 
analytic formulas in terms of orbital elements due to slow and distant encounters.
Although application of these formulae exceeds the scope of this study, alternatively 
we consider flybys in the impulsive limit.  A.P. Jackson et al. (2013, In Prep) 
provide a comprehensive and fully general set of orbital element changes due to 
impulses.  Here, however, we use simple approximations \citep{vermoe2012} and escape 
criteria to bound the effects.

\subsubsection{Impact parameter}

Within a timescale $\Delta t$, the closest expected encounter distance, 
or impact parameter $b$, is
\citep[see e.g.][]{bintre2008}
\begin{equation}
b(R) \approx \left[ 4 \pi \left( \frac{\rho_{\rm stars}(R)}{\langle
    M_{\rm stars}\rangle} \right) U(R) \Delta t \right]^{-1/2} ,
\label{ImpPara}
\end{equation}
where the quantity in the large parenthesis is the stellar number density. Here,
$U$ is the relative velocity of the stars at an infinite separation, and
$\langle M_{\rm stars} \rangle$ is the average stellar mass. In our 
calculations, we assume that in the bulge and disc, all of the mass 
is in stars, whereas in the halo, all of the mass is in dark matter.  We take 
$\langle M_{\rm stars}\rangle = 0.88M_{\odot}$ \citep{paretal2011}.

\subsubsection{Flyby velocity}

We can obtain a prescription for $U(R)$ by considering the bulge and
disc flyby stars separately. We approximate $U(R)$ by the average
relative velocity of the flyby stars to the host, so that $U(R) =
\left< (V_{\rm star} - V_{\rm fl})^2\right>^{1/2}$, where
$V_{\rm fl}$ is the velocity of the flyby star.  The value of $U(R)$ 
depends on position, as the velocities of bulge and disk stars vary 
within the Galaxy.

\begin{figure}
\centerline{
\psfig{figure=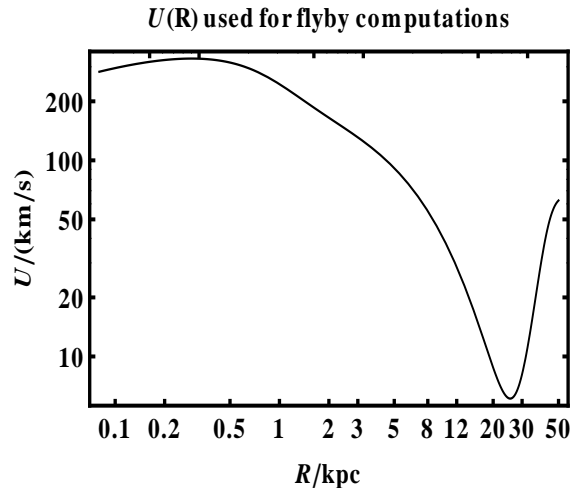,height=6.5cm,width=8cm}
}
\caption{Flyby velocity profile throughout the Milky Way for a host
  star on a circular orbit in the Galactic disc (equation
  \ref{finalVinf}).  }
\label{Flyby}
\end{figure}

The host stars are on circular orbits about the Galactic centre.
Therefore, their motion is in the azimuthal ($\phi$) direction only
with a speed given by the magnitude of the circular velocity $\vc$.
Now, let us consider the disk and bulge populations providing the
flybys. The velocity dispersion components of the (non-rotating)
Hernquist (1990) bulge are
\[
\left< V_{R, {\rm bulge}}^2 \right> 
= 
\left< V_{\theta, {\rm bulge}}^2 \right>
=
\left< V_{\phi, {\rm bulge}}^2 \right>
\]

\[
=
\frac{G M_b}{12 \epsilonW}
\bigg[
\left(
\frac{12 R \left(R + \epsilonW \right)^3}{\epsilonW^4}
\right)
\ln{\left( \frac{R+\epsilonW}{R}\right)}
.
\]

\begin{equation}
-
\left( \frac{R}{R+\epsilonW} \right)
\left(
25 + 52\frac{R}{\epsilonW} + 42 \frac{R^2}{\epsilonW^2} + 12 \frac{R^3}{\epsilonW^3}
\right)
\bigg]
.
\end{equation}
This model serves for first calculations of the effect of flybys
even though the Milky Way bulge is triaxial and rotating.

For the disc stars, the velocity dispersion around the circular speed
is taken as \citep[see e.g.][]{evacol1993,bot93,vdk11}
\begin{equation}
\left< V_{R, {\rm disc}}^2\right> = 
\frac{\left< V_{z, {\rm disc}}^2\right>}{\exp{\left(1\right)}}
=
2\left< V_{\phi, {\rm disc}}^2\right> 
=
\sigma_0^2
\exp{\left(-\frac{R}{\Rd} \right)}
,
\end{equation}
with $\sigma_0 = 100$ km s$^{-1}$. Here, we have used epicylic theory
to relate the radial and azimuthal dispersions, and the empirical
equation~(11) of \citet{vdk11} to relate the radial and vertical
dispersions.

Therefore, the square of the average relative velocity of flybys in each Galactic
component is
\[
U^2_{\rm bulge}  = \left< V_{R, {\rm bulge}}^2 \right> + \left<V_{\theta, {\rm bulge}}^2\right> + 
\left<\left(v_{\rm c} - V_{\phi, {\rm bulge}} \right)^2\right>
\]
\begin{equation}
=
\left< V_{R, {\rm bulge}}^2 \right> + \left<V_{\theta, {\rm bulge}}^2\right> + \left<V_{\phi, {\rm bulge}}^2\right>
+ v_{\rm c}^2
\end{equation}
and
\[
U_{\rm disc}^2  =
\left< V_{R, {\rm disc}}^2 \right> 
+
\left<\left(v_{\rm c} - \left(v_{\rm c} + V_{\phi, {\rm disc}} \right)
\right)^2\right> + \left< V_{z, {\rm disc}}^2 \right>
\]

\begin{equation}
= \left< V_{R, {\rm disc}}^2 \right> 
+ \left< V_{z, {\rm disc}}^2 \right>
+ \left< V_{\phi, {\rm disc}}^2\right>.
\end{equation}
Here, we have neglected the asymmetric drift and assumed that the
stellar disk rotational velocity is $v_c$.  We now weight the
contributions from bulge and disc proportionally so that
\begin{equation}
U(R) = 
U_{\rm bulge} \frac{\rho_{\rm bulge}}{\rho_{\rm
    disc} + \rho_{\rm bulge}}
+
U_{\rm disc} \frac{\rho_{\rm disc}}{\rho_{\rm
    disc} + \rho_{\rm bulge}}
\label{finalVinf}
.
\end{equation}
In Fig.~\ref{Flyby} we plot the $U(R)$ profile throughout the Milky Way.
Although it has been derived with a number of approximations and
simplifications, the profile is a useful guide to the average relative velocity
of flybys in the Galaxy.

\subsubsection{Kick velocity}

The relation between a resulting kick velocity $\Delta v$ on a
planet due to the flyby velocity $U$ is a function of the
orientation of the encounter.  We can evaluate the possible range of 
$\Delta v$ by considering the extreme cases presented by
\cite{vermoe2012}.  Let the mass of the flyby star be $M_{\rm fl}$.
The minimum and maximum $\Delta v$ are given by

\[
\Delta v = \frac{2bU^3}{G \left( M_{\rm fl} + M_{\star} \right)}
\left(1 + \frac{b^2 U^4}{G^2 \left( M_{\rm fl} + M_{\star} \right)^2}  \right)^{-2}
\]

\begin{equation}
-
\frac{2b'U^3}{G \left( M_{\rm fl} + M_{\rm p} \right)}
\left(1 + \frac{b'^2 U^4}{G^2 \left( M_{\rm fl} + M_{\rm p} \right)^2}  \right)^{-1}
,
\label{delv}
\end{equation}

\noindent{where}

\begin{equation}
b' = \left(u \pm a\right) \frac{bU^2}{G \left( M_{\rm fl} + M_{\star} \right)}
\left(1 + \frac{b^2 U^4}{G^2 \left( M_{\rm fl} + M_{\rm p} \right)^2}  \right)^{-\frac{1}{2}}
\end{equation}

\noindent{and}

\begin{equation}
u = \frac{G \left(M_{\rm fl} + M_{\star}\right)}{U^2}
\left[\sqrt{1 +  \frac{b^2 U^4}{G^2 \left( M_{\rm fl} + M_{\star} \right)^2}   } - 1 \right]
\label{ueq}
\end{equation}

\noindent{corresponding} to coplanar encounters on the opposite or same side from the planet.

\begin{figure*}
\centerline{
\psfig{figure=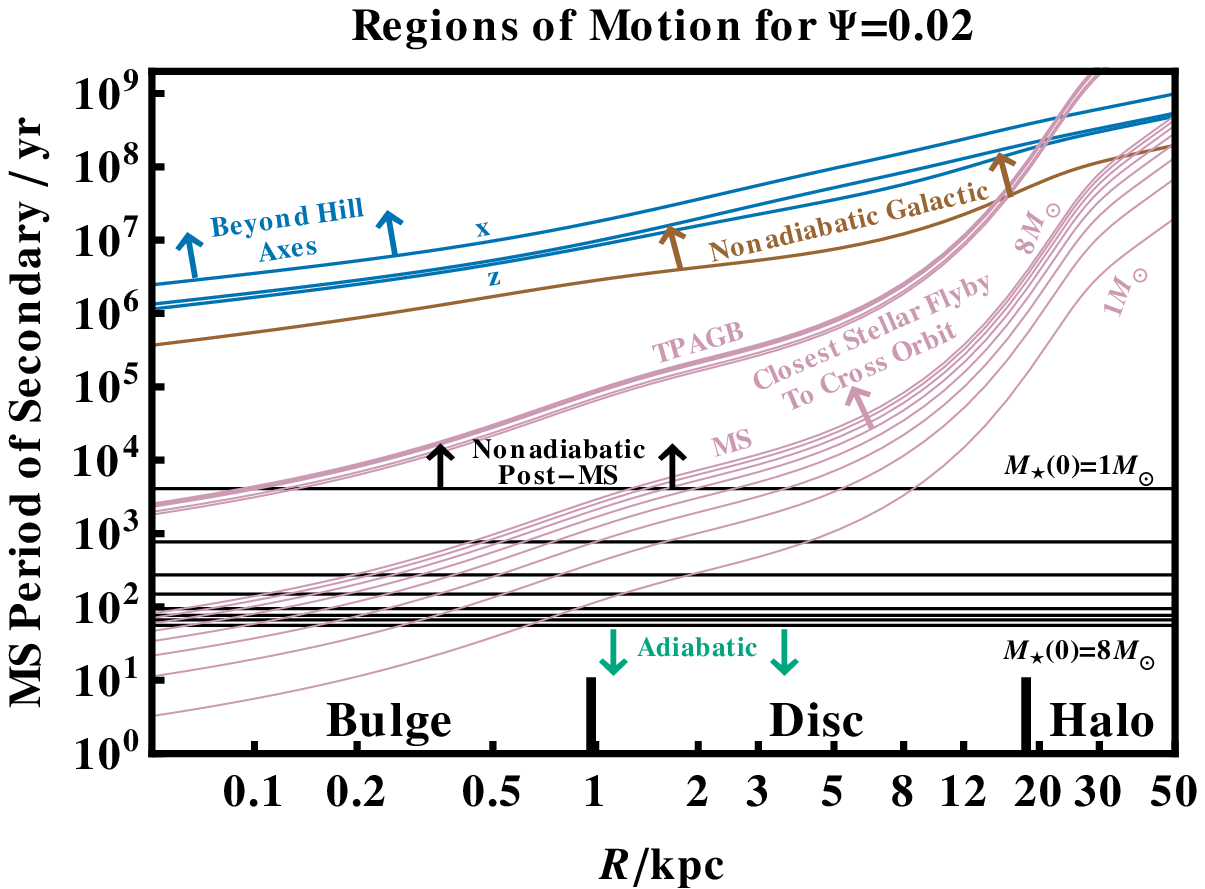,height=11.0cm,width=16.5cm} 
}
\centerline{
\psfig{figure=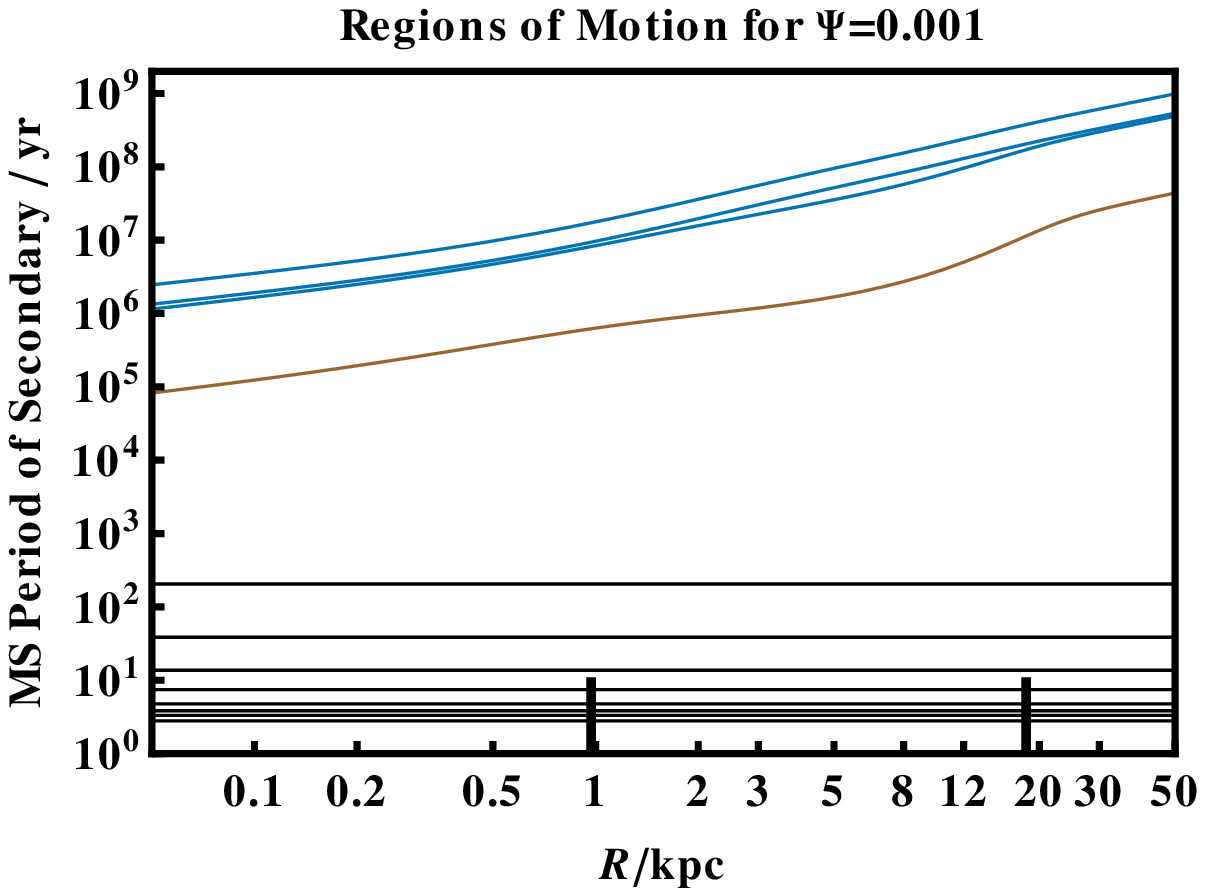,height=5.5cm,width=8.0cm} 
\psfig{figure=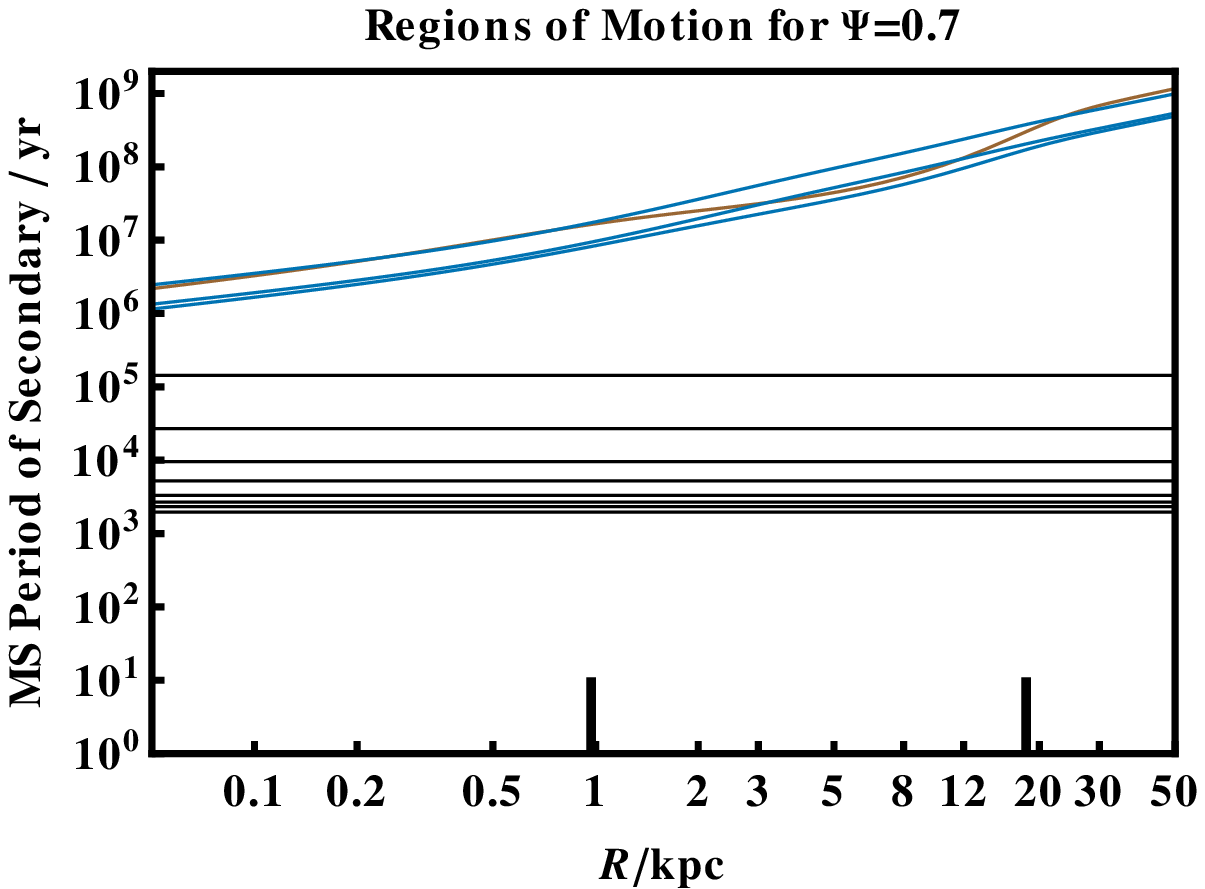,height=5.5cm,width=8.0cm} 
}
\caption{Regions where MS planets are under the influence of post-MS
stellar mass loss, Galactic tides and stellar flybys. The strength of
these first two effects are measured by the nondimensional indices
$\Psi_{\rm PMS}$ and $\Psi_{\rm MW}$, which are both set equal to $\Psi$
here.  When $\Psi$ exceeds values such as 0.001, 0.02 and 0.7, then
the system becomes nonadiabiatic to different extents.  The standard value
is approximately 0.02.  Adiabatic stellar 
mass loss increases $a$ but leaves $e$ fixed, whereas adiabatic Galactic 
tides changes $e$ but leave $a$ fixed.  Neither variable is fixed in the
nonadiabatic cases.  Close stellar flybys may or may not occur; if
they do, then their influence is strongly dependent on the
orientation of the encounter.  See Section 3.5 for a detailed
description of this figure.}
\label{MainResult1}
\end{figure*}

\subsubsection{Criteria for impulse, escape \& boundedness}

In order for the impulse approximation to be applicable, the ratio of
the orbital period of the planet to the encounter timescale must be
much greater than unity.  Hence, if we assume the closest impact
parameter is equivalent to $a$, then the criterion is equivalent to
\begin{equation}
10 \left(\frac{T}{10^5 {\rm yr}} \right)^{\frac{1}{3}} 
\left(\frac{U}{1 {\rm km \ s}^{-1}} \right) 
\left(\frac{M_{\star} + M_{\rm p}}{M_{\odot}} \right)^{-\frac{1}{3}} \gg 1
.
\end{equation}
Given that generally $U \gtrsim 10$ km s$^{-1}$ outside of stellar clusters
throughout the Milky Way's bulge and disc, the impulse approximation
should easily be applicable for orbital periods greater than tens of
years.

This approximation allows us to bound the motion of the
secondary.  The absolute minimum value of $\Delta v$
for which a planet may escape is $\Delta v_{\rm min}~=~\left(\sqrt{2}
- 1 \right) \sqrt{G \left(M_{\star} + M_{\rm p} \right)/a}$.  Similarly the
absolute maximum value of $\Delta v$ for which a planet is guaranteed
to remain bound is $\Delta v_{\rm max}~=~\left(\sqrt{2} + 1 \right)
\sqrt{G \left(M_{\star} + M_{\rm p} \right)/a}$.  With these 
criteria\footnote{See A.P. Jackson et al. (2013, In Prep) for additional information about the
effect of $\Delta v$ kicks on orbital motion.} and
equations~(\ref{delv})-(\ref{ueq}), we find that planets may either be
guaranteed to escape or guaranteed to remain bound depending entirely
on the orientation of the collision.  For example, a 
$M_{\rm p} = 0.001M_{\odot}$ planet (like Jupiter) in a circular orbit around a 
$1M_{\odot}$ star at 1000~au that is perturbed by another $1M_{\odot}$ 
star with $U = 100$~km~s$^{-1}$ and $b = 1000$~au endures a
$\Delta v$ that is anywhere between about $0.027$~km~s$^{-1}$ and $80.09$~km~s$^{-1}$, 
whereas $\Delta v_{\rm min}$~$=~0.39$ km s$^{-1}$ and 
$\Delta v_{\rm max}$~$=~2.28$ km~s$^{-1}$.  A decrease in $U$ by a factor of 10 reduces the
range of $\Delta v$ by about just an order of magnitude.  This range still
encompasses $\Delta v_{\rm min}$ and $\Delta v_{\rm max}$.  

Unfortunately then, we cannot place restrictions on how planets would
be perturbed by flybys without performing a much more detailed study.
In the absence of a more concrete measure, we can relate the closest
expected encounter distance over a given timescale (equation \ref{ImpPara}) 
with the semimajor axis of the secondary's orbit.  By equating these
two quantities, we obtain a critical semimajor axis, or orbital period, 
which will be used in the next section to compare with the effects
from mass loss and Galactic tides.  Exploring the quality of this metric, 
especially compared to the effects from multiple much more distant
stellar encounters, is a subject for future studies.

\subsection{Summary plots}

We now combine equations~(\ref{PMSindex}), (\ref{PsiMW}), (\ref{minn})
and (\ref{ImpPara}) in Fig.~\ref{MainResult1}, which is the main
result of this paper.  The plots describe different regimes of motion
due to various effects for a single planet, brown dwarf or other
companion to an evolving primary star.

The value of $\Psi$, assumed here to be equal to both $\Psi_{\rm PMS}$
and $\Psi_{\rm MW}$, indicates the extent of adiabaticity prescribed.
For example, comets with $e \approx 0.99$ need just a slight nudge to
prompt escape or collision, so that in this case one should
consider a value of $\Psi$ lower than $0.02$.

The horizontal black lines indicate the adiabatic mass-loss boundary
for stars of progenitor masses of $1 M_{\odot}$, $2 M_{\odot}$, $3
M_{\odot}$, $4 M_{\odot}$, $5 M_{\odot}$, $6 M_{\odot}$, $7 M_{\odot}$
and $8 M_{\odot}$.  We compute this boundary by identifying the
critical time $t_{\rm crit}$ at which max$[\Psi_{\rm PMS}(t)]$ equals
$0.02$, $0.7$ and $0.001$ (per each figure in the plot).  We then
derived the planet's period from $a(t_{\rm crit})$ in 
equation~(\ref{PMSindex}).  All of the stellar evolutionary tracks were
computed with {\tt SSE}.

The Galactic adiabaticity boundary (brown line) is computed with
$\Psi_{\rm MW}$ and $|\Upsilon_{zz}|$, as justified previously.  Note
importantly that the post-MS adiabaticity boundary is lower than the
Galactic adiabaticity boundary by at least an order of magnitude
always and everywhere except perhaps within a few parsecs of the
Galactic centre or for stars with masses much lower than $1M_{\odot}$.
Because any star which becomes a WD can lose at most about 80 per cent
of its mass, a planet's semimajor axis can expand adiabatically by
at most a factor of about 5.  Hence adiabatic mass loss cannot alone 
expand an orbit into the nonadiabatic Galactic region.

The region in which nonadiabatic Galactic tides act is small (about a
decade in planetary period) because the Galactic adiabaticity boundary
coincides with the Hill ellipsoid (blue line) for $\Psi_{\rm MW}
\approx 2$ (see Section 3.3).  A planet beyond the Hill ellipsoid
escapes the system.  The blue lines remain unchanged during 
the WD phase, despite the shrinkage of the Hill axis 
(equation~\ref{HillShrink}) from AGB mass loss because the 
orbital period is increased by the same factor.

The pink lines in Fig.~\ref{MainResult1} indicate  
the limit for close encounters to occur at the radius 
of the planet's orbit.  The lines were computed using
equation (\ref{ImpPara}) with $\Delta t$ representing
the Thermally Pulsing Asymptotic Giant Branch (TPAGB) 
lifetime for the top set of lines and the MS lifetime
for the bottom set of lines.  The thermally pulsing phase 
occurs near the end of giant branch evolution, when the greatest amount
of mass is lost at the greatest rate.  Any planets residing below the
top pink lines should remain undisturbed by close stellar flybys
during TPAGB evolution.  Planets evolving adiabatically due to mass
loss are generally protected from flybys.  The lower set of pink lines
indicates that close flybys may be a common, endemic feature of MS evolution.  
The flybys could play a similarly important role during WD evolution.

\begin{figure}
\centerline{
\psfig{figure=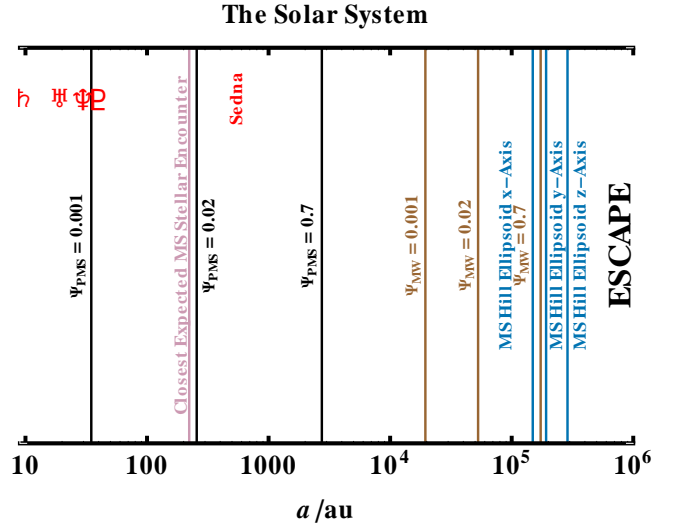,height=7.0cm,width=9.5cm}
}
\caption{Danger regions of the Solar System.  This plot
represents a vertical slice of the three plots in 
Fig. \ref{MainResult1}, assuming $M_{\star} = 1M_{\odot}$
and utilising the same colour scheme.  Lines marked
$\Psi = 0.001$ correspond to adiabaticity transitions
for objects which are sensitive to small 
changes in orbital behaviour, such as comets with 
$e \approx 0.99$.  Greater values of $\Psi$ indicate
higher thresholds for nonadiabatic orbital change.  
The locations of Saturn, Uranus, Neptune, Pluto 
(and the Kuiper Belt) and Sedna are marked in red.}
\label{SolSys}
\end{figure}

We can garner intuition for Fig.~\ref{MainResult1} as
well as help to affirm its usefulness in characterizing
individual exosystems by considering the Solar System.
Figure~\ref{SolSys} displays a vertical slice of
Fig.~\ref{MainResult1} at $R = 8$ kpc,
with $M_{\star} = 1M_{\odot}$ along the MS.
All three Hill semiaxes are plotted, as well
as the three non-adiabatic boundaries for mass
loss and Galactic tides for $\Psi = \lbrace{0.001,0.02,0.7\rbrace}$.
Additionally, the pink line corresponding to the 
limit for close encounters to occur is plotted,
and lies at about 222 au from the Sun.
This value correlates well with the 500 au, 5 Gyr Solar
neighbourhood estimate of \cite{zaktre2004} and the 
couple hundred au, full MS lifetime estimates of 
\cite{vermoe2012}.  Also, the non-adiabatic
mass loss limit for $\Psi = 0.7$, at about 2740 au,
lies within the Solar System's post-MS escape boundary range
of $10^3 - 10^4$ au assuming $\Psi = 1.0$ 
\citep{verwya2012}.  The implication
for exosystems like our Solar System is
that most secondaries which will experience
any type of non-adiabatic motion will probably also have
their orbits crossed at least once by a passing star.

\section{Characteristics of motion}

Here, we describe the types of orbital motion which are possible in
the different regions partitioned in Fig.~\ref{MainResult1}.

\subsection{Orbital elements}

Although this motion due to post-MS mass loss and Galactic tides is
represented fully by equations~(\ref{xeq})-(\ref{zeq}), we can glean a
much better understanding by converting these equations into orbital
elements.  The final result is summarized in Table 1. The variables $i$ 
$\omega$, and $\Omega$ represent the inclination, argument of pericentre,
and longitude of ascending node of the secondary orbit about the primary
with respect to a reference plane which is parallel to and coincident
with the Galactic plane.
Among the earliest derivations of the
full equations of motion due to isotropic stellar mass loss are the
studies of \cite{omarov1962} and \cite{hadjidemetriou1963}.  Later
Paper I described specific properties of these equations (in column 2
of Table 1), such as the transition from adiabaticity to
nonadiabaticity and how the pericentre of the orbit cannot ever
decrease due to stellar mass loss alone.  The orbital equations due to
Galactic perturbations were separately derived for the adiabatic
vertical case \citep[in column 3 of Table 1;][]{brasser2001}, the
adiabatic planar case \citep[in part of column 5 of Table
  1;][]{fouchard2004} and the fully nonadiabatic case \citep[in
  column 6 of Table 1;][]{vereva2013a}.

\begin{table*}
 \centering
 \begin{minipage}{180mm}
  \caption{Equations of motion in terms of orbital elements.  The
    auxiliary set of $C_i$ variables found within the table equations
    can be expressed in terms of orbital elements by
    Veras \& Evans (2013a).
\newline
$C_1 \equiv e \cos{\omega} + \cos{\left(f+\omega\right)}$, $ $ 
$C_2 \equiv e \sin{\omega} + \sin{\left(f+\omega\right)}$, $ $ 
$C_3 \equiv \cos{i} \sin{\Omega} \sin{\left(f+\omega\right)}
                   - \cos{\Omega} \cos{\left(f+\omega\right)}$, $ $ 
\newline
$C_4 \equiv \cos{i} \cos{\Omega} \sin{\left(f+\omega\right)}
                   + \sin{\Omega} \cos{\left(f+\omega\right)}$, $ $ 
$C_5 \equiv \left(3 + 4e \cos{f} + \cos{2f} \right) \sin{\omega}
        + 2 \left( e + \cos{f} \right) \cos{\omega} \sin{f}$, $ $ 
\newline
$C_6 \equiv \left(3 + 4e \cos{f} + \cos{2f} \right) \cos{\omega}
        - 2 \left( e + \cos{f} \right) \sin{\omega} \sin{f}$, $ $ 
$C_7 \equiv \left(3 + 2e \cos{f} - \cos{2f} \right) \cos{\omega}
        + \sin{\omega} \sin{2f}$, $ $ and
\newline
$C_8 \equiv \left(3  - \cos{2f} \right) \sin{\omega}
        - 2 \left(e + \cos{f} \right) \cos{\omega} \sin{f}$.
}
  \begin{tabular}{@{}ccccc@{}}
  \hline
    & Mass Loss & Mass Loss & Galactic Tides & Galactic Tides \\
 \hline
    & Adiabatic & Nonadiabatic & Adiabatic & Nonadiabatic \\
 \hline
    & Everywhere & Everywhere & Disc Only & Disc Only \\
 \hline
 $\frac{da}{dt}=$ 
& 
$   
- 
\frac{a}{\mu}\frac{d\mu}{dt} $ 
& 
$ 
- \frac{a \left(1 + e^2 + 2e \cos{f}\right)}
{1 - e^2} 
\frac{1}{\mu}\frac{d\mu}{dt}
$ 
& 
$
0
$
&
$
\frac{2a\sqrt{1-e^2}}{n\left(1+e\cos{f}\right)} 
C_1
       \left\lbrace \sin^2{i} \sin{\left(f+\omega\right)} \right\rbrace
\Upsilon_{zz}
$
 \\[5pt]
$\frac{de}{dt}=$ 
& 
$
0
$ 
& 
$
- \left(e + \cos{f} \right) 
\frac{1}{\mu}\frac{d\mu}{dt}
$
& 
$
-\frac{5e \sqrt{1 - e^2}}{2 n} \cos{\omega} \sin{\omega} \sin^2{i}
\Upsilon_{zz}
$
&
$
\frac{\left(1-e^2\right)^{\frac{3}{2}}}{2n\left(1+e\cos{f}\right)^2} 
C_6
       \left\lbrace \sin^2{i} \sin{\left(f+\omega\right)} \right\rbrace
\Upsilon_{zz}
$
 \\[5pt]
$\frac{di}{dt}=$ 
& 
$
0
$ 
& 
$
0
$
& 
$
\frac{5e^2\sin{2\omega}\sin{2i}}{8n\sqrt{1 - e^2}}
\Upsilon_{zz}
$
&
$
\frac{\left(1-e^2\right)^{\frac{3}{2}} \sin{i}}{n\left(1+e\cos{f}\right)^2} 
       \left\lbrace \cos{i} \cos{\left(f+\omega\right)} \sin{\left(f+\omega\right)} \right\rbrace
\Upsilon_{zz}
$ 
 \\[5pt]
$\frac{d\Omega}{dt}=$ 
& 
$
0
$ 
& 
$
0
$ 
& 
$
\frac{\cos{i} \left(2 + 3 e^2 - 5 e^2 \cos{2 \omega} \right)}{4n\sqrt{1 - e^2}}
\Upsilon_{zz}
$
&
$
\frac{\left(1-e^2\right)^{\frac{3}{2}}}{n\left(1+e\cos{f}\right)^2}  
       \left\lbrace \cos{i} \sin^2{\left(f+\omega\right)} \right\rbrace
\Upsilon_{zz} 
$ 
 \\[5pt]
$\frac{d\omega}{dt}=$ 
& 
$
0
$ 
& 
$
-\frac{\sin{f}}{e}
\frac{1}{\mu}\frac{d\mu}{dt}
$
& 
$
\frac{5\sin^2{\omega} \left(\sin^2i - e^2\right) - \left(1 - e^2\right)}{2n\sqrt{1-e^2}}
\Upsilon_{zz}
$
&
$
  \frac{\left(1-e^2\right)^{\frac{3}{2}}}{2en\left(1+e\cos{f}\right)^2} 
       \left\lbrace -2\sin{\left(f+\omega\right)}
  \left[ e \sin{\left(f+\omega\right)} + \frac{1}{2} C_8 \sin^2{i} \right]
       \right\rbrace
\Upsilon_{zz} 
$ 
 \\[5pt]
$\frac{df}{dt}=$ 
& 
-- 
& 
$
-\frac{d\omega}{dt}
+\frac{n \left(1 + e \cos{f}\right)^2}{\left(1 - e^2\right)^{3/2}}
$
& 
--
&
$
-\frac{d\omega}{dt}
-\cos{i} \frac{d\Omega}{dt}
+\frac{n \left(1 + e \cos{f}\right)^2}{\left(1 - e^2\right)^{3/2}}
$ 
 \\[5pt]
\hline
\end{tabular}
\end{minipage}
\end{table*}~\begin{table*}
 \centering
 \begin{minipage}{180mm}
  \setcounter{table}{0}
  \caption{Continuation}
  \begin{tabular}{cp{3cm}p{18cm}}
  \hline
    & \parbox{3cm}{\centering Galactic Tides} & \parbox{12cm}{\centering Galactic Tides} \\
 \hline
    & \parbox{3cm}{\centering Adiabatic} & \parbox{12cm}{\centering Nonadiabatic}  \\
 \hline
    & \parbox{3cm}{\centering Bulge \& Halo} & \parbox{12cm}{\centering Bulge \& Halo}  \\
 \hline
 $\frac{da}{dt}=$ 
&    
\parbox{3cm}{\centering 0}
& 
\parbox{18cm}{
$ 
  \frac{2a\sqrt{1-e^2}}{n\left(1+e\cos{f}\right)} 
\bigg[ 
\Upsilon_{zz} C_1
       \left\lbrace \sin^2{i} \sin{\left(f+\omega\right)} \right\rbrace
+
\left( \Upsilon_{xx} C_3 - \Upsilon_{xy} C_4 \right)
       \left\lbrace C_1 \sin{\Omega} \cos{i} + C_2 \cos{\Omega} \right\rbrace
$
\\
$
-
\left( \Upsilon_{yx} C_3 - \Upsilon_{yy} C_4 \right)
       \left\lbrace C_1 \cos{\Omega} \cos{i} - C_2 \sin{\Omega} \right\rbrace
\bigg]
$ 
}
 \\[5pt]
$\frac{de}{dt}=$ 
& 
\parbox{3cm}{\centering
equation (A2) of 
\\
\cite{vereva2013b} 
}
& 
\parbox{18cm}{
$
  \frac{\left(1-e^2\right)^{\frac{3}{2}}}{2n\left(1+e\cos{f}\right)^2} 
\bigg[ 
\Upsilon_{zz} C_6
       \left\lbrace \sin^2{i} \sin{\left(f+\omega\right)} \right\rbrace
+
\left( \Upsilon_{xx} C_3 - \Upsilon_{xy} C_4 \right)
       \left\lbrace C_6 \sin{\Omega} \cos{i} + C_5 \cos{\Omega} \right\rbrace
-
$
\\
$
\left(\Upsilon_{yx} C_3 - \Upsilon_{yy} C_4 \right)
       \left\lbrace C_6 \cos{\Omega} \cos{i} - C_5 \sin{\Omega} \right\rbrace
\bigg]
$
}
 \\[5pt]
$\frac{di}{dt}=$ 
& 
\parbox{3cm}{\centering
equation (A3) of 
\\
\cite{vereva2013b} 
}
& 
\parbox{18cm}{
$
  \frac{\left(1-e^2\right)^{\frac{3}{2}} \sin{i}}{n\left(1+e\cos{f}\right)^2} 
\bigg[ 
\Upsilon_{zz} 
       \left\lbrace \cos{i} \cos{\left(f+\omega\right)} \sin{\left(f+\omega\right)} \right\rbrace
-
\left( \Upsilon_{xx} C_3 - \Upsilon_{xy} C_4 \right)
       \left\lbrace \sin{\Omega} \cos{\left(f+\omega\right)} \right\rbrace
$
\\
$
+
\left( \Upsilon_{yx} C_3 - \Upsilon_{yy} C_4 \right)
       \left\lbrace \cos{\Omega} \cos{\left(f+\omega\right)}\right\rbrace
\bigg]
$ 
}
 \\[5pt]
$\frac{d\Omega}{dt}=$ 
& 
\parbox{3cm}{\centering
equation (A4) of 
\\
\cite{vereva2013b} 
}
& 
\parbox{18cm}{
$
  \frac{\left(1-e^2\right)^{\frac{3}{2}}}{n\left(1+e\cos{f}\right)^2} 
\bigg[ 
\Upsilon_{zz} 
       \left\lbrace \cos{i} \sin^2{\left(f+\omega\right)} \right\rbrace
-
\left( \Upsilon_{xx} C_3 - \Upsilon_{xy} C_4 \right)
       \left\lbrace \sin{\Omega} \sin{\left(f+\omega\right)} \right\rbrace
$
\\
$
+
\left( \Upsilon_{yx} C_3 - \Upsilon_{yy} C_4 \right)
       \left\lbrace \cos{\Omega} \sin{\left(f+\omega\right)}\right\rbrace
\bigg]
$  
}
 \\[5pt]
$\frac{d\omega}{dt}=$ 
& 
\parbox{3cm}{\centering
equation (A5) of 
\\
\cite{vereva2013b} 
}
& 
\parbox{18cm}{
$
  \frac{\left(1-e^2\right)^{\frac{3}{2}}}{2en\left(1+e\cos{f}\right)^2} 
\bigg[ 
\Upsilon_{zz} 
       \left\lbrace -2\sin{\left(f+\omega\right)}
  \left[ e \sin{\left(f+\omega\right)} + \frac{1}{2} C_8 \sin^2{i} \right]
       \right\rbrace
$
\\
$
-
\left( \Upsilon_{xx} C_3 - \Upsilon_{xy} C_4 \right)
       \left\lbrace C_8 \sin{\Omega} \cos{i} - C_7 \cos{\Omega} \right\rbrace
+
\left( \Upsilon_{yx} C_3 - \Upsilon_{yy} C_4 \right)
       \left\lbrace C_8 \cos{\Omega} \cos{i} + C_7 \sin{\Omega} \right\rbrace
\bigg]
$
}
 \\[5pt]
$\frac{df}{dt}=$ 
& 
\parbox{3cm}{\centering --}
& 
$
-\frac{d\omega}{dt}
-\cos{i} \frac{d\Omega}{dt}
+\frac{n \left(1 + e \cos{f}\right)^2}{\left(1 - e^2\right)^{3/2}}
$ 
 \\[5pt]
\hline
\end{tabular}
\end{minipage}
\end{table*}

\subsection{Escape and collision}

The properties of the equations of greatest interest here are the
ability of the companion's semimajor axis and eccentricity to expand
or shrink.  Escape can occur only if $a \rightarrow \infty$ and $e
\rightarrow 1$.  If the latter condition holds, but instead the
semimajor axis remains finite, then the companion collides with
the star.  Adiabatic mass loss causes expansion of the orbit but no change
in eccentricity, and hence can never cause a companion to escape.
Nonadiabatic mass loss also must cause expansion of the orbit,
although may cause escape because the eccentricity can be changed
(not necessarily increased).

All adiabatic Galactic tides keep a companion's semimajor axis fixed,
and hence alone can never cause escape.  However, these adiabatic
tides continuously cause a change in eccentricity, and hence can
create collisions with the parent star.  Nonadiabatic Galactic tides
can trigger both escape and collision because they may increase or
decrease $a$ or $e$.

\begin{figure*}
\centerline{
\psfig{figure=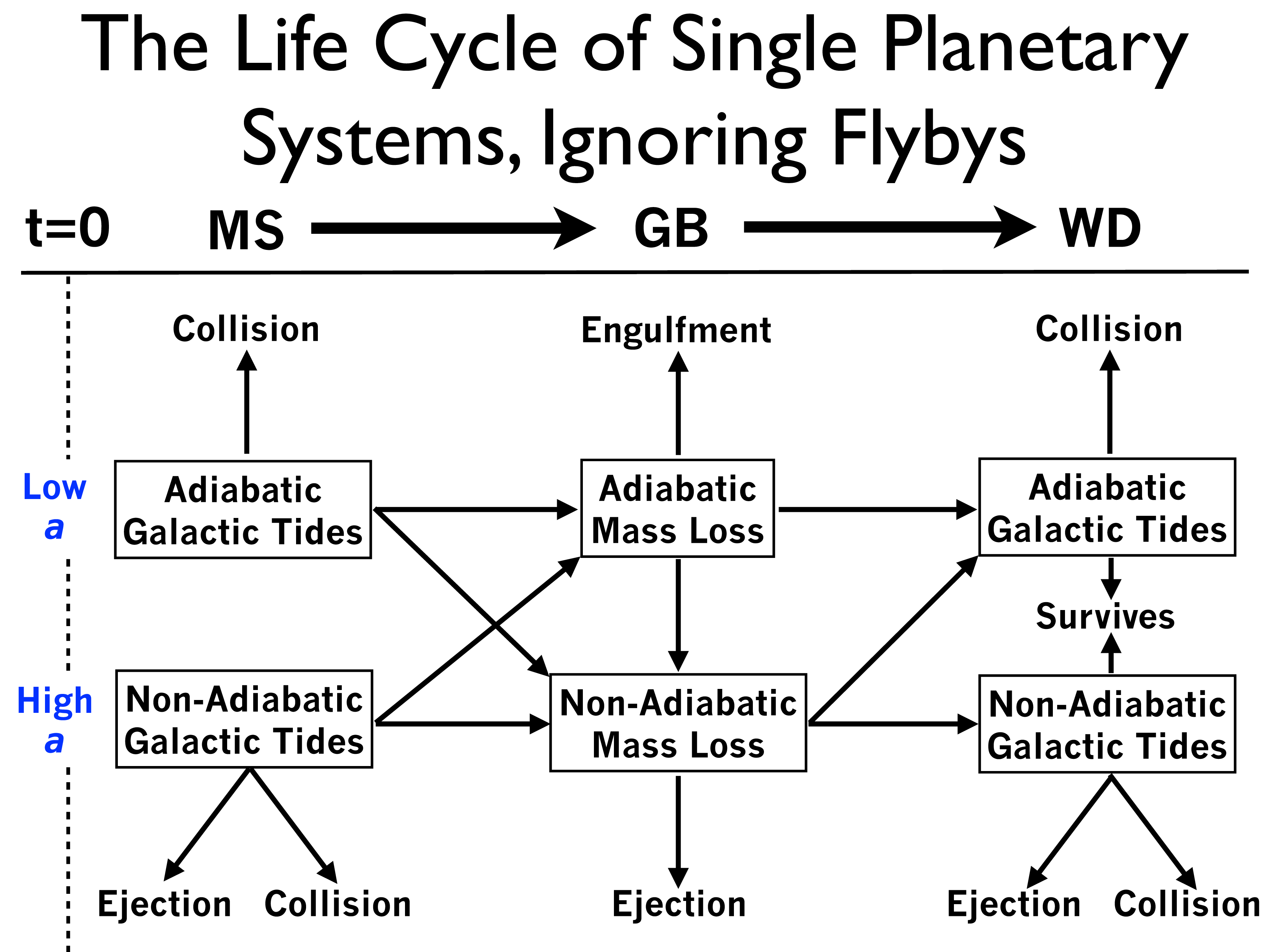,height=11.0cm,width=16.5cm} 
}
\caption{ The paths planetary lives tread for parent stars which
  eventually become white dwarfs (WDs) after a main-sequence (MS)
  phase and at least one giant branch (GB) phase.  This chart assumes 
  that planets are largely unaffected by any close flybys which may
  occur at any time.  By the phrases {\it low a} and {\it high a}, 
  we refer to semimajor axes which correspond to points below and above 
  the adiabaticity boundary lines in Fig. 3.  For example, 
  for a planet orbiting a $1 M_{\odot}$ star in the Solar neighbourhood, the semimajor 
  axis mass loss and Galactic tidal adiabaticity boundaries are at a couple 
  hundred au and several ten thousands of au.
}
\label{FlowChart}
\end{figure*}

\subsection{Summary flow chart}

Tides act continuously throughout a planet's life and post-MS
mass loss of the parent star always occurs, though just for one epoch.
In each case at every stage, instability may occur.
With these considerations, combined
with the properties of motion from Table 1 and the regions of motion
in Fig.~\ref{MainResult1}, we create the flowchart 
Fig.~\ref{FlowChart}.

The flowchart illustrates the possible evolutionary pathways of
single-planet systems, from the MS to the WD phase through the 
Giant Branch (GB) phases.  All of the mass loss occurs during
the GB phases.  The chart does not take into account flybys both
because of their uncertain occurrence rate and the variety of ways
in which they can alter the orbit of a planet or cause instability.
However, despite their exclusion, flybys can play important
roles in planetary systems, particularly for planets at wide
separations.  

The instability types listed on the bottom and top rows
include

\medskip
\noindent
1. ``collision'': when the planet collides with the star due
to Galactic tides.  In the adiabatic case, the collision occurs
because of an increase in $e$, whereas in the non-adiabatic case,
both $a$ and $e$ are varied.

\smallskip
\noindent
2. ``ejection'': when the planet escapes the system.  Both non-adiabatic
Galactic tides and non-adiabatic mass loss can prompt this action.

\smallskip
\noindent
3. ``engulfment'': when stellar tides force a planet
to spiral into the expanding envelope.
\medskip

The chart also relies on the following properties which hold true
throughout nearly the entire Milky Way. First, post-MS mass loss
during the TPAGB phase is decoupled from Galactic tides because the
latter acts on a timescale which is too slow (Fig. \ref{PMSIndex}). 
Also, adiabatic post-MS mass loss can never expand an orbit in to the nonadiabatic Galactic 
tidal regime, as also apparent from Fig.~\ref{MainResult1}. Finally, 
collision with the parent star may always occur due to Galactic tides 
during the MS or WD phases.

\subsection{Fate of adiabatic disc exoplanets}

As an application, we consider the evolutionary path that
currently-observed Galactic disc exoplanets within tens of au are
likely treading.  This path corresponds to adiabatic Galactic tidal
evolution along the MS, followed by adiabatic AGB mass loss and
concludes with adiabatic Galactic tidal evolution along the WD evolution
track.  The
majority of known and candidate planets 
(see the Extrasolar Planets Encyclopedia at http://exoplanet.eu/,
the Exoplanet Data Explorer at http://exoplanets.org/ and
the NASA Exoplanet Archive at http://exoplanetarchive.ipac.caltech.edu/) will exhibit 
this behaviour although roughly half will be destroyed by tidal 
envelope engulfment during the
GB phases.  Hence, our heavily-biased exoplanetary system observations
fail to cover the full realm of possibilities presented in
Fig.~\ref{FlowChart}.

By restricting ourselves to the disc, we can neglect all planar tidal
terms (see \citealt*{vereva2013b}). Further, by restricting ourselves
to adiabatic evolution, the relevant equations are in columns 1 and 3
of Table 1.  This set of equations does not have a general solution.

However, in special cases, exact solutions do exist for mass loss and
Galactic tides separately; for the latter, however, few results are
known.  Therefore, a coupled solution is likely to be restricted to
extensions of these known results.  Thus, we revisit the analytic
stationary solutions discussed by \cite{vereva2013b}.  In these
solutions, the planet is on a permanently polar orbit with $\omega(t)
= \omega_0 = \pm\arcsin{\left(\sqrt{1/5}\right)}$.  On this orbit,
given enough time, the planet's eccentricity becomes unity,
corresponding to instability.  Slight deviations from these exact
conditions cause substantial perturbations (change of several
tenths in eccentricity) over a MS lifetime (about 10 Gyr for
a $1 M_{\odot}$ star) for wide-orbit companions (at a few thousand au)
in the Galactic disc.

For the purposes of the below computation, assume that mass is being
lost at a constant rate such that $\dot{\mu} = G \dot{M}_{\star}
\equiv -\kappa$, where $\kappa > 0$.  We attempted to
solve the coupled differential equations, now including mass loss, 
in the same Taylor-expanded manner\footnote{The Taylor expansion
is carried out to fourth-order in eccentricity about $e=0$ and first-order in
inclination about $i = 90^{\circ}$.} as \cite{vereva2013b}.
However, we obtained no physical solutions.  Instead, we found
that the exact stationary orbit case ($i = 90^{\circ}$) does yield solutions, both of 
which can provide insights into the interplay between mass
loss and the Galactic tide.  The solutions are:

\begin{equation}
e(t) = 
\frac{e_0}
{
\cosh{\left[ 
\frac{t \mu_0 \Upsilon_{zz}}
{n_0 \left(\kappa t - \mu_0 \right)} 
\right]}
\mp
\sqrt{1 - e_{0}^2}
\sinh{\left[ 
\frac{t \mu_0 \Upsilon_{zz}}
{n_0 \left(\kappa t - \mu_0 \right)} 
\right]}
}
.
\end{equation}

\noindent{The} eccentricity increases to unity,
given enough time.  The corresponding survival time is

\[
t_{\rm surv}
\approx
\frac{n_0}
{ \frac{n_0}{\mu_0} \kappa \mp 2 \Upsilon_{zz}
\left[ \ln{\left( \frac{2-e_{0}^2}{e_{0}^2} 
                + \frac{2\sqrt{1 - e_{0}^2}}{e_{0}^2}
\right)    }  \right]^{-1}
     }
\]

\begin{equation}
=
\frac{1}
{n_0
\left(
\Psi_{\rm PMS}(0) \mp \Psi_{\rm MW}(0) \left[ \ln{\left( \sqrt{\frac{2-e_{0}^2}{e_{0}^2} 
+ \frac{2\sqrt{1 - e_{0}^2}}{e_{0}^2}}\right)  }  \right]^{-1}
\right)
}
.
\label{survTime}
\end{equation}
With the exception of the mass loss term and the additional 
term in the large radicand, this expression is 
equivalent to Equation (36) of \cite{vereva2013b}. 
 
The denominator of equation (\ref{survTime}) illustrates the 
interplay between stellar mass loss and Galactic tides.  
For initially circular orbits, the survival time
is dictated entirely by post-MS evolution\footnote{One cannot,
  however, reduce this expression by assuming $\Upsilon_{zz}
  \rightarrow 0$ because in this case $e(t) = e_0$ and $t_{\rm surv}$ has
  no physical meaning.}.  As the planet's eccentricity approaches
unity, while its semimajor axis is increasing, potential outcomes
include colliding with the star or leaving the adiabatic regime.
Escape through the Hill sphere, however, is not possible through this
mechanism.

To make better sense of this equation, consider a planet in the Solar 
neighbourhood on a $10^2$ yr orbit (adiabatically evolving with mass loss 
and Galactic tides) around a $2 M_{\odot}$ star such that $n_0 = 2\pi/10^2$yr.  
From Fig. \ref{PMSIndex}, we have $\Psi_{\rm PMS} < 10^{-3}$ and 
$\Psi_{\rm MW} \approx 10^{-12}$.  The quantity in square brackets is a typically 
order-of-unity weak function of $e_0$, varying for example between 0.3 and 3.0 for 
values of $e_0$ approximately equal to $0.9565$ and $0.0994$.  Therefore, in this system the mass loss 
dominates, so in fact $t_{\rm surv} \approx \mu(0)/\dot{\mu}(0)$, which is the 
time taken for the star to lose all its mass.  Because stars that become WDs lose 
at most about 80 per cent of their mass, $t_{\rm surv}$ is never reached.

As we demonstrate in Fig.~\ref{PMSIndex}, for orbital periods
corresponding to adiabatic mass loss, $\Psi_{\rm PMS} \gg \Psi_{\rm
  MW}$ always.  For these two terms to be comparable, the logarithmic
coefficient of $\Psi_{\rm MW}(0)$ (containing $e_0$) must be so high
that in this case the secondary would collide with the parent star.
This example demonstrates that, aside from exceptional circumstances,
Galactic tides play a negligible role during post-MS mass loss 
(as opposed to during MS or WD evolution).

\section{Discussion}

We now consider the implications of this work on Oort clouds, stellar
binary companions and free-floating planets.

\subsection{Depletion of Oort clouds after AGB mass loss}

MS Oort cloud comets close to the Hill ellipsoid boundary are
susceptible to escape during AGB mass loss as the Hill ellipsoid
shrinks and the orbit expands.  The Hill shrinkage scales as
$M_{\star}^{1/3}$ (equation~\ref{HillShrink}) but the orbital
expansion is more complex because of the dependence on eccentricity
(column 2 of Table 1).  The eccentricity variation prevents us from
effectively bounding the nonadiabatic semimajor axis variation except
in specific cases (Paper I).  A lower bound to the semimajor axis
expansion roughly approximates the adiabatic limit\footnote{For
  example, Fig.~4 of Paper I helps demonstrate that sub-adiabatic
  semimajor axis expansion may occur just during the adiabatic
  transition region.  In real systems, this transition region is
  unlikely to persist for the majority of GB evolution.}.  If we use
this approximation, then the semimajor axis expansion scales as
$M_{\star}^{-1}$.

\begin{figure}
\centerline{
\psfig{figure=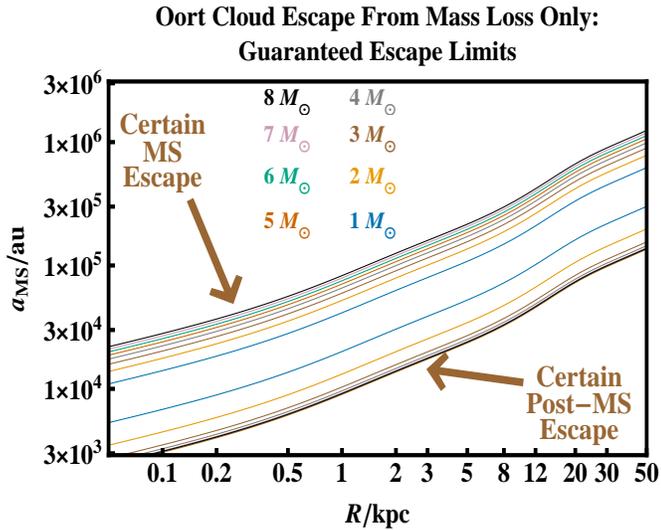,height=7cm,width=9cm} 
}
\caption{Boundaries beyond which Oort cloud escape
definitely occurs on the main-sequence (MS) (upper 8
curves) and during post-MS evolution (lower 8 curves).
Each similarly-coloured pair of curves corresponds
to a different stellar progenitor mass, which are labeled.
In every case, the MS escape boundary is over twice
as far away as the post-MS escape boundary.
The plot demonstrates that Oort clouds become
depleted and excited due to post-MS mass loss.}
\label{CritOort}
\end{figure}

We combine these two scalings to produce Fig.~\ref{CritOort}, in which
we plot Oort cloud semimajor axes beyond which the comets must escape
on the MS (upper 8 curves) and during post-MS evolution due to mass 
loss alone (lower 8 curves).  Each pair of coloured curves corresponds
to a different progenitor stellar mass.  The distance between the upper
and lower curves in each pair increases for higher stellar masses because
those exhibit greater mass loss.  Each curve corresponds to the
$z$ semiaxis of the Hill ellipsoid, which is always the smallest of the
three semiaxes; a comet will not be able to remain bound
for an entire orbit if part of that orbit is outside of the Hill ellipsoid.

Flybys, which are not included in this
estimate, will still likely play a role in Oort cloud evolution
during TPAGB mass loss (Fig.~\ref{MainResult1}).  Nevertheless, flybys
can keep comets bound only in a fraction of cases.  
Figure~\ref{CritOort} suggests, for example, that for $1 M_{\odot}$ stars in the Solar
Neighbourhood, comets beyond about $10^5$ au would not remain bound
during AGB mass loss.  This result corresponds well to fig.~8 of
\cite{braetal2010}.  Oort clouds in the Galactic bulge are
particularly vulnerable to escape.  They can remain bound only if they
extend inward to semimajor axes under a few thousand au.  This
limitation suggests that the bulge might be teaming with free-floating
comets.

The post-MS danger region is substantial.  The fraction of the entire MS Hill ellipsoid
that contains the semimajor axis range corresponding to guaranteed mass 
loss-induced escape is approximately equal to 
$1$~$-$~$\left[1 - M_{\star}(t_{\rm WD})/M_{\star}(t_{\rm MS}) \right]^{4/3}$, or 
51.5\%, 74.7\%, 82.7\%, 85.3\%, 87.3\%, 87.8\%, 88.3\% and 89.0\%
for progenitor stellar masses of $1M_{\odot}$, $2M_{\odot}$, $3M_{\odot}$, 
$4M_{\odot}$, $5M_{\odot}$, $6M_{\odot}$, $7M_{\odot}$, and $8M_{\odot}$.
The location of Oort cloud comets at the onset of post-MS mass loss
then crucially determines how populated the Milky Way will become 
with free-floating comets as stars die.

Comets which do remain bound are likely to be highly dynamically
excited in eccentricity \citep[e.g.][]{verwya2012} and inclination.
In the latter case, any anisotropy in the mass loss has a 
pronounced effect in the outer reaches of planetary systems 
\citep{paralc1998}, and specifically grows in importance as $\sqrt{a}$
and affects the inclination evolution \citep{veretal2013a}.
Consequently, the flux of eccentric comets entering the inner WD
system might change significantly and could have
direct or indirect consequences for explanations of the origin
of polluted WD atmospheres
\citep[e.g.][]{zucetal2003,ganetal2008,faretal2009}.

\subsection{Two evolving stars}

In a binary system with two stars, evolutionary possibilities become
significantly more numerous.  While both are on the MS,
Fig.~\ref{MainResult1} still applies, except that the nonadiabatic
limits are being shifted by order unity due to the secondary mass being of
the same order as a Solar mass in equations~(\ref{PMSindex}-\ref{PsiMW}).  
The onset of post-MS evolution can
elicit many possible outcomes \citep{huretal2002} depending on the
stars' masses, metallicities and mutual orbit.  If the stars are
sufficiently separated from each other that they evolve independently,
then the system might undergo two distinct phases of MS evolution
before both stars become WDs.  If both stars lose mass concurrently,
the evolution of the mutual orbit is still dictated by the equations
in column 2 of Table 1.  Nevertheless, in this case the nonadiabatic
limits in Fig.~\ref{MainResult1} might be shifted more significantly.

We can investigate concurrent stellar evolution in more detail, 
particularly to determine in what cases the secondary moves off of the MS 
during post-MS evolution of primary.  The most significant coupling of mass 
loss rates would occur for mass loss on the AGB, and so we consider only that 
phase.  We set primary progenitor masses of $2M_{\odot}$, $3M_{\odot}$, $4M_{\odot}$, 
$5M_{\odot}$, $6M_{\odot}$, $7M_{\odot}$ and $8M_{\odot}$, and assume the primaries and 
secondaries are born at the same time, evolve independently, and initially contain 
Solar metallicity. We then use {\tt SSE} to trace the evolution of lower-mass 
secondaries, in decrements of $0.02 M_{\odot}$ from the initial primary mass, until 
the total mass loss of the system is always dictated by just one of the stars.
All simulations make the same assumptions about mass loss as in the rest of the
paper.  By using {\tt SSE} 
rather than the binary version, {\tt BSE} \citep{huretal2002}, we avoid introducing many additional 
parameters which are unnecessary for this basic exploration.

We find that in nearly every case, the secondary mass must be within 10 per 
cent of the progenitor primary mass in order to achieve mass loss coupling during 
their AGB phases.  Specifically, we find that the highest secondary progenitor masses 
that demonstrate decoupling are $1.82, 2.74, 3.70, 4.68, 5.62, 6.56, 7.16 M_{\odot}$.  
The probability that a binary pair is so closely matched in mass varies as a function of 
stellar type.  O-type binaries typically feature nearly equal-mass stars \citep{zinyor2007} 
whereas less massive stars demonstrate a greater range of mass ratios 
\citep[e.g.][]{methil2009}.  In this scenario, mass loss 
coupling amongst the highest mass binaries will have the greatest
consequences for shifting the 
nonadiabatic limits in Fig. \ref{MainResult1}. However, recent work claiming that the mass ratio 
distribution is essentially flat for all primaries more massive than $0.3 M_{\odot}$ 
\citep{duckra2013} demonstrates that perhaps the coupling probability is largely 
independent of total system mass.

A planet may be introduced in this system orbiting one of the stars
(circumstellar) or orbiting both stars (circumbinary).  Tens of
percent of all confirmed exoplanets reside in binary systems.  Most of
these planets orbit in a circumstellar manner.  However, recent
circumbinary exoplanet discoveries
\citep{doyetal2011,oroetal2012a,oroetal2012b,weletal2012} have helped
incite interest in multiple stellar-component planetary systems.

We do not fully calculate systems with a third body but we can provide 
a rough estimate of the
consequences.  The circumbinary case may be reduced to a 2-body system
if the binary separation is much less than the distance to the
orbiting planet.  In paper II we demonstrated that, in this case, the key
quantity to determine the orbital evolution is the overall mass-loss
rate from the binary system.  Paper II also showcased how this rate is a
complex function of the phase space.  Applied to our
Fig.~\ref{MainResult1}, the position of each black horizontal line
would be a function of the properties of both stars.  The other figure
curves would remain largely unchanged.  In paper II we demonstrated that
circumbinary systems feature, generally, much smaller mass-loss
adiabaticity boundaries than for the single-star case.  Consequently,
the potential evolutionary pathways in Fig.~\ref{FlowChart} would
remain unchanged.

For circumstellar orbits in binary systems, planetary motion can no
longer be modelled with Figs.~\ref{MainResult1} and \ref{FlowChart}.
Even without any mass loss, planets may bounce between binary
stars on the MS \citep{moever2012}.  Post-MS evolution can then
cause an exchange reaction allowing a planet to hop from the
evolving star to the non-evolving companion, around which the planet can
attain a stable orbit \citep{kraper2012}.  How Galactic tides
and stellar flybys alter these and similar situations represents
an interesting topic for future investigation.

\subsection{Free-floating bodies}

This work's emphasis on escape motivates discussion of the Milky Way's
population of substellar-mass objects which are not gravitationally
bound to a single or small group of stars.  These bodies have been
referred to in the literature by several terms, including
free-floating planets and cluster planets (see e.g. the Extrasolar Planet
Encyclopedia at http://exoplanet.eu/), rogue planets, sub brown-dwarfs
\citep[e.g. Pg. 7 of][]{perryman2011}, nomads
\cite[e.g.][]{stretal2012} and isolated planetary-mass objects
\citep[e.g.][]{deletal2012}.  The ambiguity arises from both the
uncertain deuterium-burning mass limit \citep[see e.g.][]{spietal2011}
and the multiple potential formation pathways for these objects.

Observations of free-floaters have been plagued by uncertain mass
determinations that arise from a degeneracy between the luminosity,
mass and age of the objects.  This uncertainty has placed doubt as to
whether any of these objects are below the deuterium-burning mass
limit, despite the mass errors quoted (see the Extrasolar Planet
Encyclopedia).  Recently, this degeneracy has finally been broken:
\cite{deletal2012} has discovered and associated the $4M_J$-$7M_J$
T-type exoplanet CFBDSIRJ214947.2-040308.9 with the AB Doradus moving
group with an 87\% probability.  Membership helps ensure that this
free-floater is of planetary-mass and ultimately that free-floating
planets do exist.

This near-confirmation, as well as the astonishing estimate that the
Milky Way's free-floating giant planet population exceeds the bound
giant planet population \citep{sumetal2011}, motivates determination
of the free-floating planet formation pathways.  Although molecular
cloud collapse triggers star formation, the resulting fragmentation
may also produce planetary-mass objects which become free-floating.
Alternatively, free-floating planets may have been originally formed
from protoplanetary disc fragmentation, and only later in life have
escaped due to instability within the system.  Many candidate
free-floating planets have been discovered in young, star-forming
clusters (see references on Pg. 215 of \citealt*{perryman2011}) and in
the field (see references in \citealt*{deletal2012}), thereby largely
failing to constrain the dominant formation pathway.  Free-floaters
formed by cloud fragmentation could later appear isolated in the
field, and free-floaters formed by disc fragmentation could be
stripped off of their parent stars in both clusters
\citep{adaetal2006,freetal2006,spuetal2009,maletal2011,boletal2012,parqua2012}
and in the field \citep{zaktre2004,varetal2012,vermoe2012}.
Additional sources include ejection from planet-planet scattering
\citep[e.g.][]{verray2012} and post-main sequence instability (Paper
I; Paper II; \citealt*{adetal2013}; \citealt*{musetal2013a,musetal2013b}; 
\citealt*{veretal2013b}; \citealt*{voyetal2013}).

Our work suggests that if a population of planets exist
on the MS at distances corresponding to $a \gg 100$ au,
then the combined effects of post-MS mass loss,
Galactic tides and flybys yields a free-floating planet population
that may be strongly dependent on age and Galactic location.  The
oldest regions of the bulge would feature the highest sub-populations,
and the youngest regions of the halo the lowest.  The overall
population is unlikely to scale strongly with stellar mass, because
the adiabatic mass loss limits in all three plots of
Fig.~\ref{MainResult1} are well-separated from the adiabatic Galactic
tidal limit.  Only beyond this limit is escape possible (see
columns 3-6 of Table 1) after a planet survives post-MS evolution and
withstands the effects of stellar flybys.  The distribution of Oort
cloud comets in the interstellar medium is similarly dependent on
age and location.  However, these objects are not yet detectable.  The
forthcoming all-sky survey missions GAIA and LSST will be able to
detect free-floaters which have masses that exceed about 1~$M_J$,
whereas WFIRST will be able to probe the population of free-floaters
at lower masses \citep{stretal2012}.

\section{Conclusion}

We have identified when and where planetary, cometary and binary
systems are most affected by the collusion of three important
perturbations, post-MS mass loss, Galactic tides and stellar flybys.
Two of these forces are external, and one internal; two of these must
occur, whereas one may occur.  By considering the entire range of WD
progenitor stars in the Galactic bulge, disc and halo with all
possible primary-secondary separations (Fig.~\ref{MainResult1}), we
obtain a representative picture of the possible evolutionary pathways
and potential instabilities experienced by a stellar system with a
single substellar companion (Fig.~\ref{FlowChart}).

Despite a complete analytical characterization of the orbital changes
induced by two of these effects (Table~1), the unknown prevalence and
orientation of stellar flybys, and their open-ended consequences
(Section 3.5.4) prevent us from establishing conclusions more specific
than the ones listed below.  An outstanding feature of nearly all
these conclusions is their applicability throughout the Milky Way.

\smallskip
\noindent
1. {\bf Everywhere in Galaxy:} AGB mass loss is decoupled from Galactic tides
(Fig.~\ref{PMSIndex}): the former dominates over the duration of the
mass loss. Also, barring exceptional circumstances, AGB mass loss 
will run its course without disruption from close stellar flybys if the
secondary is close enough to the primary such that the mass loss is
adiabatic.

\smallskip
\noindent
2. {\bf Everywhere in Galaxy:} Adiabatic mass loss cannot 
by itself expand the secondary's orbit
into the nonadiabatic Galactic region.

\smallskip
\noindent
3.  {\bf Everywhere in Galaxy:} Adiabatic Galactic tidal effects cannot cause escape and the
smallest Hill axis is always a factor of about $(2/\Psi_{\rm
  MW})^{3/4} \approx 2-300$ beyond the nonadiabatic Galactic boundary.  Hence, if a
secondary survives AGB mass loss but fails to reach this boundary, the
body can escape the WD only through flybys.

\smallskip
\noindent
4.  {\bf Everywhere in Galaxy:} The combined effects of Hill ellipsoid
shrinkage and secondary orbital excitement due to AGB mass loss
(Fig.~\ref{CritOort}) are likely to decimate a bound MS Oort cloud
population and alter the cometary flux into the inner WD system.

\smallskip
\noindent
5.  {\bf Everywhere in Galaxy:} The evolution of binary stars can also 
be approximated as in Figs.~\ref{MainResult1} and \ref{FlowChart}
unless they are close enough to each other to tidally interact, and
only if they leave the MS at different times.

\smallskip
\noindent
6.  {\bf In bulge only:} All types of secondaries 
except for hot Jupiters or similarly-tight binary companions
endure strong perturbations from close flybys.
Consequently, we expect the frequency of ejections,
and generally the amount of free-floating material, to be
significantly more abundant in the bulge than in the disc
or halo.

\section*{Acknowledgments}

We thank the referee for useful comments that
have improved the quality of this manuscript.
This work benefited from support by the European Union
through ERC grant numbers 279973 and 320964.  CAT thanks 
Churchill College for his fellowship.

\label{lastpage}

\end{document}